\documentstyle[]{mn}
\footnotesize
\newdimen\minuswidth    
\setbox0=\hbox{$-$}
\minuswidth=\wd0
\catcode`@=\active
\def@{\kern\minuswidth}
\newdimen\digitwidth    
\setbox0=\hbox{\rm0}
\digitwidth=\wd0
\catcode`!=\active
\def!{\kern\digitwidth}
\normalsize

\title[Timing pulsars in 47~Tuc]{Timing the millisecond pulsars in 47~Tucanae}
\author[P. C. Freire et al.]{
P. C. Freire$^1$,
F. Camilo$^{2}$,
D. R. Lorimer$^3$,
A. G. Lyne$^1$,
R. N. Manchester$^4$
\newauthor
and N. D'Amico$^5$\\
$^1$University of Manchester, Jodrell Bank Observatory, Macclesfield,
Cheshire, SK11~9DL, UK\\
$^2$Columbia Astrophysics Laboratory, Columbia University,
550~West~120th~Street, New York, NY~10027, USA\\
$^3$NAIC, Arecibo Observatory, HC3~Box~53995, Arecibo, PR~00612, USA\\
$^4$Australia Telescope National Facility, CSIRO, P.O.~Box~76, Epping,
NSW~1710, Australia\\
$^5$Osservatorio Astronomico di Bologna, Via Ranzani 1, 40127 Bologna,
Italy
}

\date{\scriptsize Accepted for publication by MNRAS, 2001 March 9}
\begin{document}

\maketitle
\newcommand{\setthebls}{
}

\setthebls

\begin{abstract}
In the last ten years 20 millisecond pulsars have been discovered in
the globular cluster 47~Tucanae. Hitherto, only three of these pulsars
had published timing solutions. Here we improve upon these three and
present 12 new solutions. These measurements can be used to determine a
variety of physical properties of the pulsars and of the cluster.  The
15 pulsars have positions determined with typical errors of only a few
milliarcseconds and they are all located within $1\farcm2$ of the
cluster centre.  Their spatial density within that region is consistent
with a distribution of the type $n(r) \propto r^{-2}$, with a sudden
cutoff outside 4 core radii.  Two pulsars have a projected separation
of only $0\farcs12$, and could be part of a triple system containing
two observable pulsars.  We have measured the proper motions of five of
the pulsars: the weighted mean of these, $\mu_{\alpha} = (6.6 \pm
1.9)$\,mas\,yr$^{-1}$ and $\mu_{\delta} = (-3.4 \pm
0.6)$\,mas\,yr$^{-1}$, is in agreement with the proper motion of
47~Tucanae based on {\em Hipparcos\/} satellite data.  The period
derivatives measured for many of the pulsars are dominated by the
dynamical effects of the cluster gravitational field, and are used to
constrain the surface mass density of the cluster.  The pulsar
accelerations inferred from the observed period derivatives are
consistent with those predicted by a King model using accepted cluster
parameters.  We derive limits on intrinsic pulsar parameters: all the
pulsars have characteristic ages greater than 170\,Myr and have
magnetic fields smaller than $2.4\times10^9$\,Gauss; their average
characteristic age is greater than $\sim 1$\,Gyr.  We have also
measured the rate of advance of periastron for the binary pulsar
J0024$-$7204H, $\dot{\omega} = (0.059 \pm 0.012)^{\circ}$\,yr$^{-1}$,
implying a total system mass $1.4^{+0.9}_{-0.8}\,{\rm M}_{\odot}$ with
95\% confidence.

\end{abstract}


\section{Introduction}
\label{sec:intro}

\begin{table*}
\begin{center}
\caption{Parameters of the globular cluster 47~Tuc.}
\begin{tabular}{ l l l}
\hline
Parameter & Value & Reference \\
\hline
Centre R.A., $\alpha_{\rm 47~Tuc}$  (J2000) \dotfill
& $\rm 00^{\rm h} 24^{\rm m} 05\fs29 \pm 0\fs28$ &\cite{dps+96} \\
Centre Decl., $\delta_{\rm 47~Tuc}$ (J2000) \dotfill
& $\rm -72^\circ04'52\farcs3 \pm 1\farcs3$ &\cite{dps+96}\\
Distance, $D$ \dotfill 
& $\rm 5.0 \pm 0.4\,kpc$ &\cite{rei98b} \\
Age \dotfill 
& $\rm 10.0 \pm 0.4\,Gyr$ &\cite{gfc+97} \\
Total mass \dotfill 
& $(1.07 \pm 0.04)\times 10^{6}\,{\rm M}_{\odot}$  & \cite{mey89} \\
Tidal radius \dotfill 
& $40'$ (58\,pc)  &\cite{dac79} \\
Core radius, $\theta_c$ ($r_c$) \dotfill 
& $\rm 23\farcs1 \pm 1\farcs7$ (0.6\,pc) & \cite{hgg00} \\
Escape velocity \dotfill 
& $\rm 58\,km\,s^{-1}$ & \cite{web85} \\
Central density \dotfill
& $\sim 1 \times 10^5 \rm M_{\odot} \rm pc^{-3}$ & \cite{pm93} \\
Central line-of-sight velocity dispersion, $v_{z}(0)$ \dotfill
& $\rm 11.6 \pm 1.4 \,km\,s^{-1}$ &\cite{mm86b} \\
Proper motion in R.A., $\mu_{\alpha}$ \dotfill
& $\rm 7.0 \pm 1.0\,mas\,yr^{-1}$  & \cite{obgt97}\\
Proper motion in Decl., $\mu_{\delta}$ \dotfill
& $\rm -5.3 \pm 1.0\,mas\,yr^{-1}$ & \cite{obgt97}\\
\hline
\end{tabular}

\label{tab:47tuc}
\end{center}
\end{table*}

Among the globular clusters in the Galactic system, 47~Tucanae
(henceforth 47~Tuc) holds the record for the number of known pulsars:
20 to date \cite{mlr+91,rlm+95,clf+00}.  Camilo et al. estimated the
total number of potentially detectable pulsars in this cluster to be at
least 200.  The known pulsar population of 47~Tuc is very different
from the population in the Galactic disk: all of the pulsars  have
periods less than 8\,ms, and 13 are members of binary systems. Radio
images of the cluster \cite{fg00,ma00} show three scintillating point
sources, the positions of which coincide with the brightest known pulsars in
the cluster:  47~Tuc~C, D and J. It is therefore improbable that any
bright pulsars remain to be discovered, even with the severe selection
effects against the detection of binaries with very short orbital
periods and pulsars with very short rotational periods noted by Camilo
et al.~(2000).

In addition to this set of pulsars, there is a collection of other
exotic objects near the core of 47~Tuc: at least 9 X-ray sources
(Hasinger, Johnston \& Verbunt 1994; Verbunt \& Hasinger 1998)
\nocite{hjv94,vh98} and more than 20 blue stragglers \cite{gysb92}.
Evidence has also been found for large numbers of cataclysmic variables
in the core (Edmonds et al. 2001, in preparation). The high stellar
density necessary for the formation of these objects may increase the
rate of exchange of stars in binaries \cite{hmv91}, leading to the
formation of low-mass X-ray binaries and millisecond pulsars, a process
addressed by Rasio, Pfahl \& Rappaport (2000)\nocite{rpr00} for the
specific case of 47~Tuc.

Table~\ref{tab:47tuc} summarises the basic parameters of 47~Tuc which
we use to interpret our results. This compilation has five important
updates on the information tabulated by Camilo et al.~(2000). First,
the position of the centre of 47~Tuc is taken to be point ``D'' of
plate 1 of \nocite{dps+96} De Marchi et al.~(1996).  We derive the
absolute position of ``D'' by comparing its position with that of point
``G'' on the same plate, whose absolute position was obtained by
Guhathakurta et al.~(1992).  Second, recent {\em Hipparcos\/} data
\cite{rei98b} yields a distance to 47~Tuc of 5.0\,kpc, 10\% larger than
previously thought \cite{web85}.  Third, according to Howell,
Guhathakurta \& Gilliland~(2000)\nocite{hgg00}, the core radius is
double the value used previously \cite{dps+96}.  In addition, the
stellar line-of-sight velocity dispersion at the centre \cite{mm86b} is
smaller than the previously accepted value of 13.2\,km s$^{-1}$
\cite{web85}.  The fifth update is the proper motion, obtained from
{\em Hipparcos\/} data: $\mu_{\alpha} = (7.0 \pm 1.0)$\,mas\,yr$^{-1}$
and $\mu_{\delta} = (-5.3 \pm 1.0)$\,mas\,yr$^{-1}$ \cite{obgt97}.

The evolutionary state of the cluster is not known precisely. It is
generally presumed (e.g., Djorgovski \& King 1984\nocite{dk84}) that
the core is in a pre-collapse phase. However, other authors (e.g., de
Marchi et al. 1996) suggest that the cluster underwent core collapse in
the past, reaching a core radius similar to that of M15, and that it
has now re-expanded with the energy provided by the creation and
hardening of stellar binary systems, an important process in globular
cluster evolution \cite{spi87}. This hypothesis was based on a
determination for the angular core radius of $12''$, which indicated a
dynamically evolved cluster; and it can explain the large number of
exotic objects observed in the cluster. However, the more recent
determination of core radius \cite{hgg00} doubles this value, casting
some doubt on that conclusion, since the re-expansion would have had to
be far more vigorous.  Therefore the evolutionary state of the cluster
remains unclear. In \S~\ref{sec:binaries} we discuss the properties of
the observed binary pulsars, and their relevance for the solution of
this problem is addressed in \S~\ref{sec:concl}.

Prior to the present work, the scarcity of coherent timing solutions
for the pulsars in 47~Tuc did not permit the studies that require a
large sample of well-timed pulsars in a single cluster to be carried
out.  This has only been possible to date for the eight pulsars in M15
(Anderson 1992; Phinney 1992, 1993). These studies of M15 provide much
of the theoretical background for the present work and allow a useful
comparison between M15 and 47~Tuc.

The new observations presented here bring the number of pulsars with
coherent timing solutions known in 47~Tuc to 15. In \S~\ref{sec:obs} we
describe the observations and data reduction and analysis used to
obtain the new timing solutions.  We present these solutions in
\S~\ref{sec:solutions}.  The new solutions lead to a wealth of
astrophysical results which we discuss in the rest of the paper.  In
\S~\ref{sec:positions} we use high-precision astrometry to investigate
the pulsar distribution in 47~Tuc and the proper motion of the
cluster.  In \S~\ref{sec:accel} we derive some constraints for the
surface mass density of the cluster, and limits on the characteristic
ages and magnetic fields of the pulsars.  In \S~\ref{sec:binaries} we
discuss some characteristics of the binary systems, and present the
rate of advance of periastron measured for 47~Tuc~H.  Finally, in
\S~\ref{sec:concl}, we summarise our results and briefly discuss the
prospects for future timing observations of the pulsars in 47~Tuc.

\section{Observations and data processing}
\label{sec:obs}

Most of the early observations of 47~Tuc were made at 660\,MHz using
the 64-m radio telescope at Parkes, Australia, between 1989 and 1991
\nocite{mld+90,mlr+91} (Manchester et al.~1990; Manchester et al.~1991;
for a summary of all observations see Table~\ref{tab:log}). These
observations led to the discovery of 10 millisecond pulsars. Further
observations of 47~Tuc at 430\,MHz between 1991 and 1993 \cite{rlm+95}
resulted in the discovery of one more pulsar,
B0021$-$72N\footnote{Because all pulsars are detected in the same
telescope beam pattern, and precise positions were not available
originally, the pulsars were named B0021$-$72A, B, etc.  Objects A, B
and K were later found to be spurious, so the list of pulsars in 47~Tuc
starts with B0021$-$72C, now referred to as J0023$-$7204C, or
47~Tuc~C.}, and phase-coherent timing solutions for 47~Tuc~C and D, the
two brightest isolated pulsars in the cluster. For the remaining nine
pulsars then known, the paucity of detections prevented the
determination of any further coherent timing solutions.

\begin{table*}
\begin{center}
\caption{Summary of 10 years of observations of 47~Tuc.  $N_{\rm obs}$
is the number of observing days, $\nu_{c}$ is the central radio
frequency used, $\Delta \nu$ is the total bandwidth recorded, $N_{\rm
chan}$ is the number of frequency channels per polarization, $\tau$ is
the most common integration time, $t_{\rm samp}$ is the sampling time,
and $t_{\rm res}$ is the time resolution for each set of observations
at the central frequency.  With $\Delta \nu_{\rm chan} \equiv \Delta
\nu / N_{\rm chan}$, the time resolution is obtained using $t_{\rm
res}^{2} = t_{\rm DM}^{2} + t_{\rm samp}^{2}$, where $t_{\rm DM} = 8.3
\times (\rm DM/\rm cm^{-3} \rm pc) \times (\Delta \nu_{chan}/MHz)
\times (\nu_{c}/GHz)^{-3}\,\mu$s. }
\begin{tabular}{ r r r r r r r r }
\hline
\multicolumn{1}{c}{Time span}      &
\multicolumn{1}{c}{$N_{\rm obs}$}  &
\multicolumn{1}{c}{$\nu_{c}$}      &
\multicolumn{1}{c}{$\Delta \nu$}   &
\multicolumn{1}{c}{$N_{\rm chan}$} & 
\multicolumn{1}{c}{$\tau$}         &
\multicolumn{1}{c}{$t_{\rm samp}$} &
\multicolumn{1}{c}{$t_{\rm res}$}  \\
 &  &\multicolumn{1}{c}{(MHz)}   &
\multicolumn{1}{c}{(MHz)}        &    &
\multicolumn{1}{c}{(min)}        &
\multicolumn{1}{c}{($\mu\rm s$)} &
\multicolumn{1}{c}{($\mu\rm s$)} \\
\hline
7/89--5/92  &  47 &  660 &  32 & 128 &  74    & 300 & 350 \\
5/91--11/93 &  42 &  430 &  32 & 256 &  90/60 & 300 & 440 \\
1/98--4/98  &  18 &  660 &  32 & 256 & 105    & 125 & 155 \\
8/97--8/99  & 108 & 1374 & 288 &  96 & 280    & 125 & 270 \\
\hline
\end{tabular}

\label{tab:log}
\end{center}
\end{table*}

In August 1997, observations of 47~Tuc were resumed at Parkes after a
four-year gap. The cluster was observed on 126 days during the
following 24 months. The majority of these observations were made using
the central beam of a multi-beam system \cite{lcm+00} at a central
frequency of 1374\,MHz ($\lambda 20$\,cm) with a bandwidth of
288\,MHz.  Incoming signals from two orthogonal polarisations were
down-converted and filtered in a $2\times96\times3$-MHz filter bank.
After summing the polarisation pairs, the resulting voltages were
one-bit sampled every $125\,\mu$s and, together with accurate time
referencing, stored on magnetic tapes for later analysis.

Recording the raw data in this way has two benefits. First, it allows
off-line searches for new pulsars to be carried out.  This strategy has
been remarkably successful, essentially doubling the number of pulsars
known, many of which are undetectable most of the time because of
interstellar scintillation \cite{clf+00}.  Second, for timing purposes,
it allows us to refine the timing models iteratively, resulting in the
ephemerides reported here.

In the timing analysis, the raw data are de-dispersed and folded
according to an initial ephemeris for each pulsar. In the early stages,
this ephemeris is determined from variations in the observed rotational
period of the pulsar in the discovery and confirmation observations.
It is important that the pulsars be detected frequently, otherwise it
is impossible to count unambiguously the number of pulsar rotations
between two different epochs. The 20-cm observations made after August
1997 made use of the excellent sensitivity of the Parkes multi-beam
system and resulted in a large increase in the detection rate for all
the previously known pulsars in 47~Tuc when compared to the earlier
observations at 660\,MHz. This was despite the lower flux densities at
1400\,MHz.

The integrated pulse profiles obtained by folding the data at the
predicted pulse period are then cross-correlated with a low-noise
``standard'' pulse profile (see Camilo et al. 2000 for the 20-cm
profiles).  This allows the determination of topocentric pulse
times-of-arrival (TOAs), referred to the observatory time standard.
This was related to UTC(NIST) by a radio link to the Tidbinbilla Deep
Space Station and from there to NIST by a GPS common-view system. We
then use the {\sc tempo} software
package\footnote{http://pulsar.princeton.edu/tempo} to calculate the
corresponding barycentric TOAs using the assumed pulsar position and
the JPL DE200 solar system ephemeris (http://ssd.jpl.nasa.gov).  The
differences between the measured and predicted TOAs are used to improve
the parameters of the ephemeris (for further details of this process
see, e.g., Taylor 1992\nocite{tay92}). In the early stages, the
improved ephemeris is used to reprocess the raw data.  This iterative
process increases the number and quality of the TOAs, which in turn are
used to improve the ephemeris.

After determining the timing solutions for the 1997--1999 period, we
re-analysed the raw data from the earlier observations (see
Table~\ref{tab:log}). This resulted in a substantial increase in the
number and quality of TOAs, especially for pulsars in binary systems,
because of the much improved orbital ephemerides.

\begin{table}
\begin{center}
\begin{scriptsize}
\caption{Timing status for the pulsars in 47~Tuc. J2000 names have not
been assigned to pulsars for which there is presently no timing
solution. A horizontal line separates the pulsars discovered in the
early 1990s \protect\cite{rlm+95} from those reported by Camilo et
al.~(2000). $N_{\rm TOA}$ is the number of TOAs used in each
solution, and rms is the weighted root-mean-square of the post-fit
timing residuals in $\mu$s. }
\begin{tabular}{ l c c c r r }
\hline
Pulsar                                         &
\multicolumn{1}{c}{Orbital}                    &
\multicolumn{4}{c} {Coherent timing solutions} \\
(J2000)                           &
\multicolumn{1}{c}{solution}      &
\multicolumn{1}{c}{Start MJD}     &
\multicolumn{1}{c}{Final MJD}     &
\multicolumn{1}{c}{$N_{\rm TOA}$} &
\multicolumn{1}{c}{rms}           \\
\hline
0023$-$7204C & isolated  & 48494 & 51406 & 750 & 19 \\
0024$-$7204D & isolated  & 48492 & 51405 & 402 & 11 \\
0024$-$7205E & known     & 48464 & 51406 & 404 & 22 \\
0024$-$7204F & isolated  & 48491 & 51406 & 348 & 14 \\
0024$-$7204G & isolated  & 48600 & 51404 &  72 & 26 \\
0024$-$7204H & known     & 48517 & 51406 & 299 & 27 \\
0024$-$7204I & known     & 50683 & 51406 & 119 & 27 \\
0023$-$7203J & known     & 48491 & 51406 &1114 &  7 \\
0024$-$7204L & isolated  & 50686 & 51406 &  59 & 39 \\
0023$-$7205M & isolated  & 48517 & 51406 &  80 & 50 \\
0024$-$7204N & isolated  & 48515 & 51405 &  69 & 14 \\
\hline
0024$-$7204O & known     & 50683 & 51406 & 210 & 20 \\
47~Tuc~P     & known     & \dots & \dots & \dots & \dots \\
0024$-$7204Q & known     & 50689 & 51405 & 146 & 29 \\
47~Tuc~R     & known     & \dots & \dots & \dots & \dots \\
47~Tuc~S     & known     & \dots & \dots & \dots & \dots \\
0024$-$7204T & known     & 50683 & 51380 & 125 & 63 \\
0024$-$7203U & known     & 48515 & 51383 & 178 & 20 \\
47~Tuc~V     & not known & \dots & \dots & \dots & \dots \\
47~Tuc~W     & known     & \dots & \dots & \dots & \dots \\
\hline
\end{tabular}

\label{tab:solutions}
\end{scriptsize}
\end{center}
\end{table}

\section{Coherent timing solutions for 15 pulsars}
\label{sec:solutions}

\begin{table*}
\begin{center}
\caption{Celestial coordinates and rotational parameters for 15 pulsars
in 47~Tuc at the reference epoch of MJD~51000. In this and in the
following tables, the numbers in parentheses are 1-$\sigma$
confidence-level uncertainties in the last digits quoted.  The
positions are referred to the J2000 equinox. For all pulsars, these
parameters were fitted assuming that the pulsar has the proper motion
of the cluster given in Table~\protect\ref{tab:47tuc} (see
\S~\protect\ref{sec:pm}). For the DMs, the error is one or less in the
last digit quoted. }
\begin{tabular}{ c l l l l c }
\hline
Pulsar &
\multicolumn{1}{c}{Right Ascension} &
\multicolumn{1}{c}{Declination}     &
\multicolumn{1}{c}{Period}          &
\multicolumn{1}{c}{$\dot{P}$}       &
\multicolumn{1}{c}{DM} \\
       &
\multicolumn{1}{c}{($^{\rm h}\,\,\,^{\rm m}\,\,\,^{\rm s}$)} &
\multicolumn{1}{c}{($^\circ\,\,\,' \,\,\,''$)}    &
\multicolumn{1}{c}{(ms)}                                  &
\multicolumn{1}{c}{($10^{-20}$)}                          &
\multicolumn{1}{c}{($\rm cm^{-3}\,pc$)} \\
\hline
C & 00 23 50.3511(2)  & $-$72 04 31.486(1)  & 5.7567799980968(5) &
!$-$4.985(2)   & 24.6 \\
D & 00 24 13.8776(2)  & $-$72 04 43.8323(9) & 5.3575732850382(5) &
!$-$0.333(2)   & 24.7 \\
E & 00 24 11.1013(4)  & $-$72 05 20.131(3)  & 3.5363291476529(7) &
!@9.852(2)     & 24.2 \\
F & 00 24 03.8519(2)  & $-$72 04 42.799(1)  & 2.6235793491667(3) &
!@6.451(1)    & 24.4 \\
G & 00 24 07.956(1)   & $-$72 04 39.683(7)  & 4.040379145748(2)  &
!$-$4.215(5)   & 24.4 \\
\multicolumn{6}{c}{}  \\
H & 00 24 06.6989(7)  & $-$72 04 06.789(3)  & 3.210340709441(1)  &
!$-$0.162(5)   & 24.4 \\
I & 00 24 07.932(1)   & $-$72 04 39.664(5)  & 3.484992064038(1)  &
!$-$4.59(1)   & 24.4 \\
J & 00 23 59.4040(1) & $-$72 03 58.7720(9) & 2.1006335458586(2)  &
!$-$0.9787(4) & 24.6 \\
L & 00 24 03.770(2)   & $-$72 04 56.90(1)   & 4.346168005784(5)  &
$-$12.19(3)    & 24.4 \\
M & 00 23 54.485(3)   & $-$72 05 30.72(1)   & 3.676643219590(3)  &
!$-$3.832(6)    & 24.4 \\
\multicolumn{6}{c}{}  \\
N & 00 24 09.1835(9)  & $-$72 04 28.875(7)  & 3.053954347392(1)  &
!$-$2.186(2)   & 24.6 \\
O & 00 24 04.6492(5)  & $-$72 04 53.751(3)  & 2.6433432956679(7) &
!@3.032(6)     & 24.4 \\
Q & 00 24 16.488(1)   & $-$72 04 25.149(7)  & 4.033181182805(2)  &
!@3.41(2)     & 24.3 \\
T & 00 24 08.541(2)   & $-$72 04 38.91(2)   & 7.588479792133(9)  &
@29.47(6)      & 24.4 \\
U & 00 24 09.8325(5)  & $-$72 03 59.667(3)  & 4.342826691451(2)  &
!@9.524(3)     & 24.3 \\
\hline
\end{tabular}

\label{tab:common_parameters}
\end{center}
\end{table*}

\begin{table*}
\begin{center}
\begin{small}
\caption{Orbital parameters for the eight binary pulsars in 47~Tuc with
known coherent timing solutions.  For 47~Tuc~H the value listed under
$T_{\rm asc}$ is the time of passage through periastron. }
\begin{tabular}{ c l l l l l l }
\multicolumn{7}{c}{}\\
\hline
Pulsar &
\multicolumn{1}{c}{$P_{b}$}       &
\multicolumn{1}{c}{$x$}           &
\multicolumn{1}{c}{$T_{\rm asc}$} &
\multicolumn{1}{c}{$\omega$}      &
\multicolumn{1}{c}{$e$}           &
\multicolumn{1}{c}{$\dot{\omega}$}\\
 & 
\multicolumn{1}{c}{(days)}        &
\multicolumn{1}{c}{(sec)}         &
\multicolumn{1}{c}{(MJD)}         &
\multicolumn{1}{c}{($^\circ$)}    & &
\multicolumn{1}{c}{($^\circ\,\mbox{yr}^{-1}$)}\\
\hline
E & 2.256844818(6) & 1.981839(4) & 51000.4194595(8) & 218.5(7)   & 0.000318(4) &\dots \\
H & 2.3576965(5)   & 2.152812(7) & 51000.97359(6)   & 110.494(9) & 0.070557(5) & 0.059(12)\\
I & 0.229792249(8) & 0.038458(9) & 51000.014865(8)  & \dots      & $<0.001$    &\dots \\
J & 0.1206649386(2)& 0.0404087(6) & 51000.0416882(3) & \dots     & $<0.0003$    &\dots \\
O & 0.135974304(2) & 0.045157(5) & 51000.021155(2)  & \dots      & $<0.0004$   &\dots \\
Q & 1.18908405(1)  & 1.462200(7) & 51000.985847(2)  & 132(9)     & 0.00008(1) &\dots \\
T & 1.12617678(2)  & 1.33847(3)  & 51000.317048(3)  & !55(6)     & 0.00037(4)  &\dots \\
U & 0.4291056829(6)& 0.526953(5) & 51000.0705011(6) & 341(7)     & 0.00015(2)  &\dots \\
\hline
\end{tabular}

\label{tab:binaries}
\end{small}
\end{center}
\end{table*}

The coherent timing solutions obtained from the analysis described in
the previous section result in a wealth of high-precision astrometric,
spin and (for the binary pulsars) orbital information.
Table~\ref{tab:solutions} summarises the current timing status for the
pulsars known in 47~Tuc. The orbital parameters for 47~Tuc~S and T were
first determined using a new technique described by Freire, Kramer \&
Lyne (2001)\nocite{fkl01}, while the orbital parameters of the
remaining binaries had been determined earlier (Robinson et al. 1995;
Camilo et al. 2000).  The timing solutions were obtained as follows.

Because the dispersion in proper motions among pulsars is expected to
be very small (see \S~\ref{sec:pm}) we assume that all pulsars have the
average proper motion of the cluster, whose value we take to be that
determined from {\em Hipparcos} data (Table~\ref{tab:47tuc}).  Using
{\sc tempo}, the available 1991--1999 TOAs were fitted to a model for each
pulsar containing celestial coordinates, spin parameters, and binary
elements where relevant.  Dispersion measures (DM) were obtained
separately by measuring frequency-dependent delays across the 288-MHz
bandwidth available at 1400\,MHz, and the resulting values coincide
with those given by Camilo et al. (2000).  We also fitted
astronomically meaningless time offsets between groups of TOAs obtained
at different frequencies and with different time resolution.  We do
this because it is difficult to make a proper absolute alignment of
pulse profiles obtained at different frequencies owing to variations in
pulse shapes.

\begin{figure*}
\setlength{\unitlength}{1in}
\begin{picture}(0,9)
\put(-4.5,0){\includegraphics{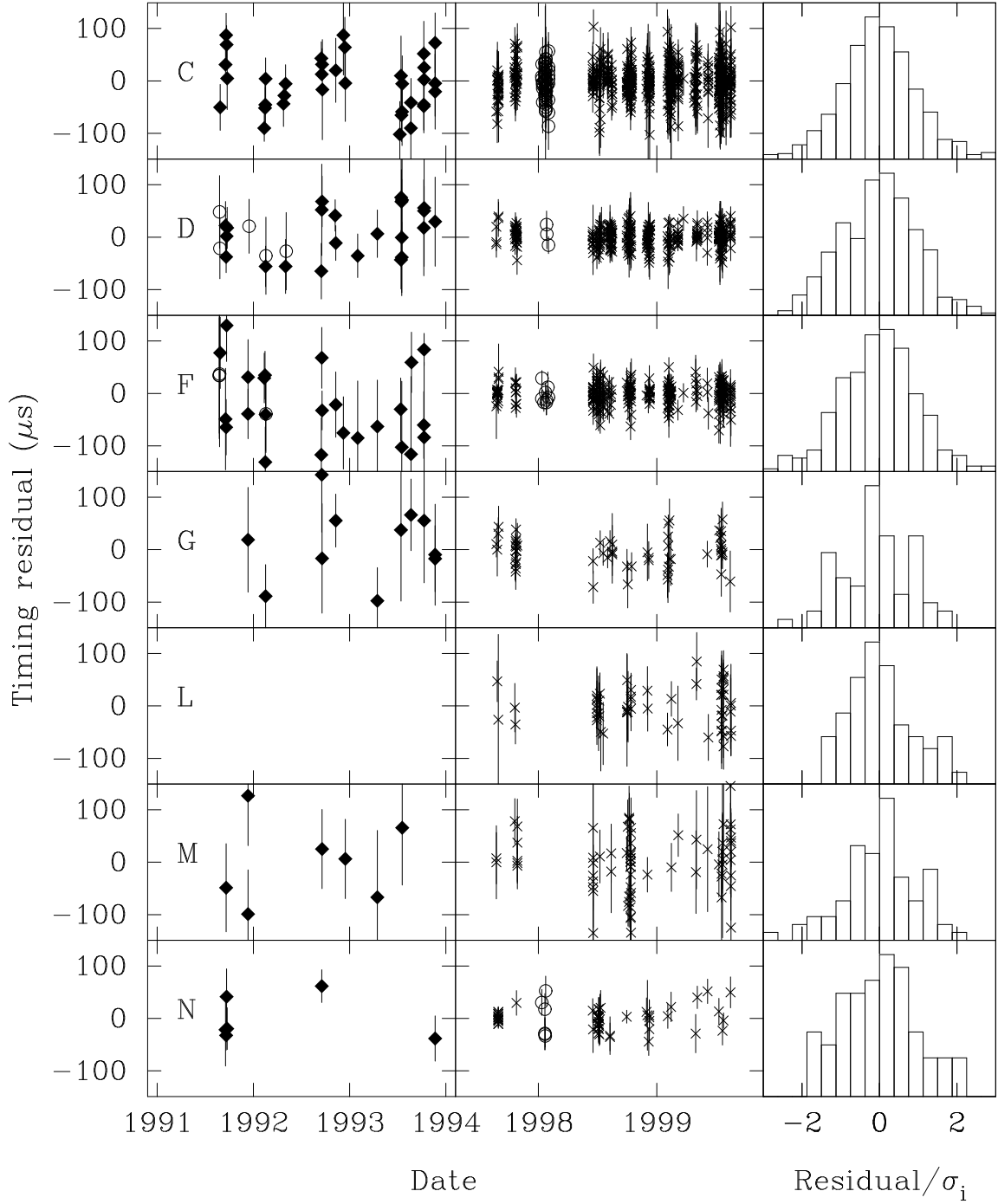}}
\end{picture}
\caption [] {Post-fit timing residuals as a function of date for seven
isolated pulsars. In this and in Figs.~\ref{fig:resmjd2} and
\ref{fig:resorb} the crosses, circles and filled diamonds represent
residuals at 1400, 660 and 430\,MHz respectively. At right we display
corresponding histograms of timing residual/uncertainty for individual
TOAs. }

\label{fig:resmjd1}
\end{figure*}

In Table~\ref{tab:common_parameters} we present the positions,
rotational parameters, and DMs obtained in this way for 15 pulsars.
The corresponding timing residuals for the seven isolated pulsars are
displayed in Fig.~\ref{fig:resmjd1} as a function of time.

\begin{figure*}
\setlength{\unitlength}{1in}
\begin{picture}(0,9.0)
\put(-4.5,-0.5){\includegraphics{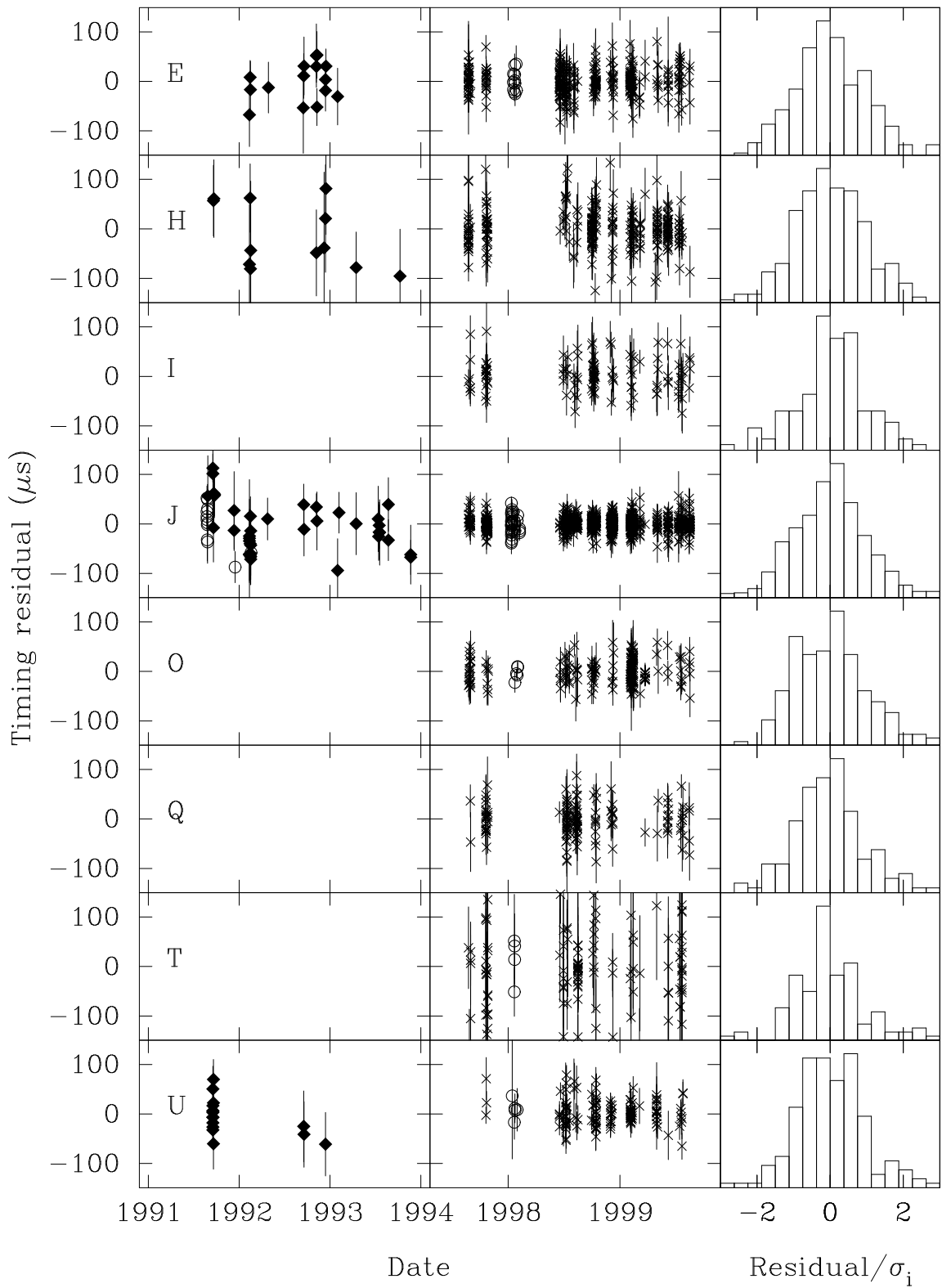}}
\end{picture}
\caption [] {Post-fit timing residuals as a function of date for eight
binary pulsars and corresponding histograms of timing
residual/uncertainty for individual TOAs. }

\label{fig:resmjd2}
\end{figure*}

\begin{figure*}
\setlength{\unitlength}{1in}
\begin{picture}(0,9.0)
\put(-4.5,-0.5){\includegraphics{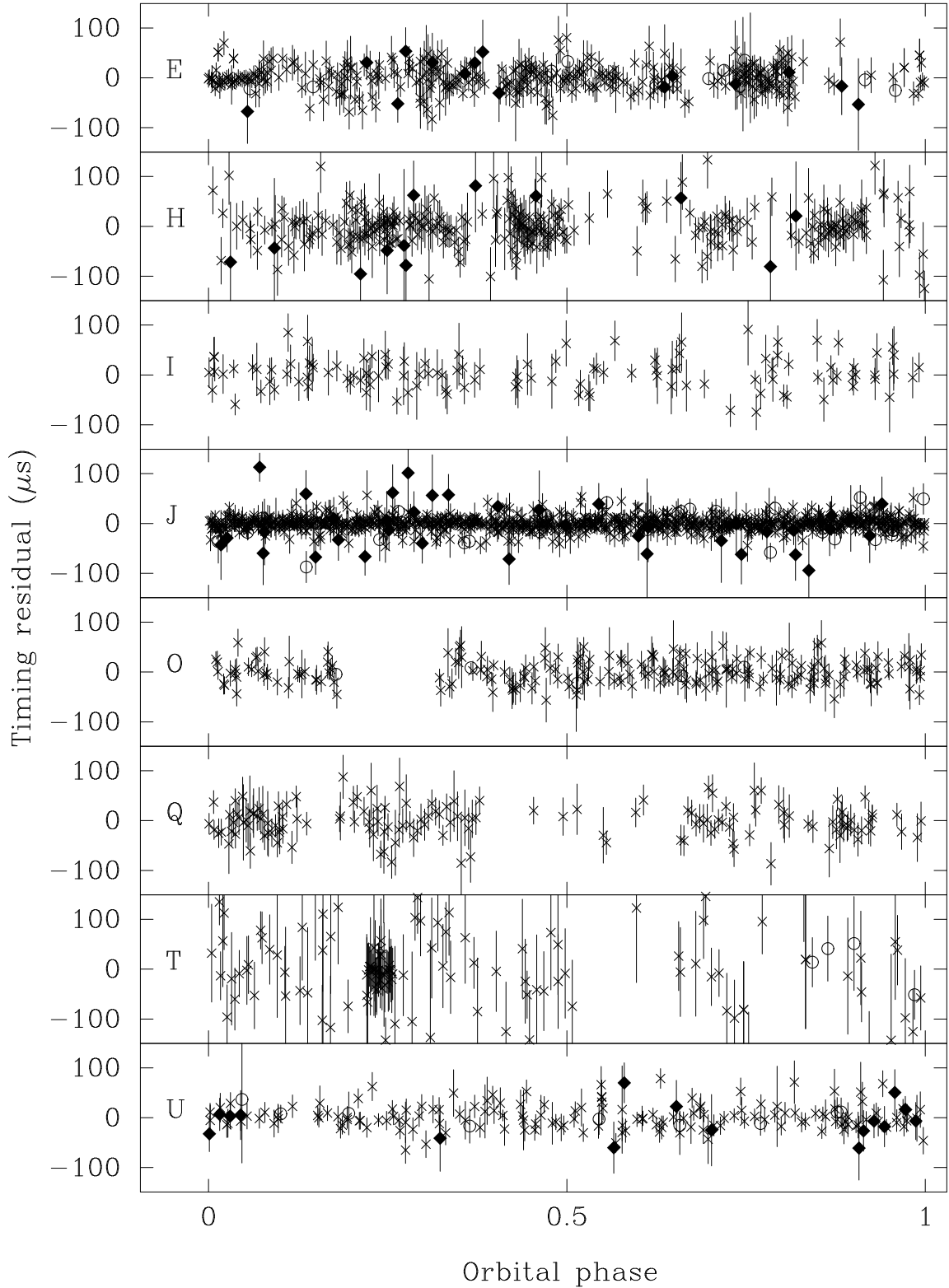}}
\end{picture}
\caption [] {Post-fit timing residuals as a function of orbital phase
for the eight binary pulsars whose residuals are shown as a function
of date in Fig.~\ref{fig:resmjd2}.  Phase is measured relative to
periastron for 47~Tuc~H and to ascending node for the remaining
pulsars.  47~Tuc~O is eclipsed for a portion of its orbit, and we
do not include TOAs from this region in the timing fit. }

\label{fig:resorb}
\end{figure*}

The orbital elements for the eight binary pulsars with timing solutions
are presented in Table~\ref{tab:binaries}, while the corresponding
timing residuals are shown in  Fig.~\ref{fig:resmjd2} as a function of
time, and in Fig.~\ref{fig:resorb} as a function of orbital phase.

For each pulsar we list the five Keplerian parameters: binary period
$P_b$, projected semi-major axis light travel time $x$, time of passage
through the ascending node $T_{\rm asc}$ and, where measurable,
longitude of periastron $\omega$ and eccentricity $e$.  For 47~Tuc~H we
indicate the time of passage through periastron, rather than through
ascending node, and we also list the measured rate of advance of
periastron $\dot \omega$.  For low-eccentricity binaries (all but
47~Tuc~H), in which some orbital parameters are highly covariant in
standard binary fits, we used the binary model ELL1 as implemented in
{\sc tempo} to determine the solutions presented (see Lange et
al.~2001).  \nocite{lcw+00}

Finally we attempted to measure proper motions for those pulsars where
we had a combination of a long time baseline and generally high-quality
data.  We did this with a straightforward {\sc tempo} fit, and list the
results in Table~\ref{tab:proper_motions} for five pulsars.  We also
determined proper motions by comparing the positions of the pulsars in
1992--1993 and 1998--1999: for each of the two independent position
measurements, no proper motion was assumed, and all TOAs used at each
epoch had the same frequency (430\,MHz in early 1990s, 1400\,MHz in
late 1990s), thus avoiding any problems with the alignment of standard
profiles at different frequencies.  In these fits we used the accurate
rotational and binary parameters obtained previously, so only the
celestial coordinates were determined.  A proper motion was then
calculated from the differences in position between the two epochs.
These measurements are consistent with those made in the global fits,
but they are less precise.

All uncertainties presented in
Tables~\ref{tab:common_parameters}--\ref{tab:proper_motions} are our
best estimates of realistic 1-$\sigma$ confidence levels, and in most
cases are twice the formal fit uncertainties from {\sc tempo}.

\begin{table}
\begin{center}
\caption{Proper motions of five pulsars in 47~Tuc. The average is
obtained by weighting the proper motions by the inverse of each
uncertainty. }
\begin{tabular}{ c l l }
\hline
Pulsar & \multicolumn{1}{c}{$\mu_{\alpha}$}
       & \multicolumn{1}{c}{$\mu_{\delta}$}  \\
       & \multicolumn{1}{c}{(mas yr$^{-1}$)}
       & \multicolumn{1}{c}{(mas yr$^{-1}$)}  \\
\hline
C & 7.8(1.6)         & $-$3.2(1.0) \\
D & 7.7(1.8)         & $-$2.6(1.6) \\
E & 9.1(2.3)         & $-$3.9(2.1) \\
F & 6.6(2.6)         & $-$3.7(2.0) \\
J & 4.6(0.8)         & $-$3.7(0.8) \\
 & & \\
Average: & 6.6(1.9)  & $-$3.4(0.6) \\
\hline
\end{tabular}

\label{tab:proper_motions}
\end{center}
\end{table}

The main factors limiting the precision of the timing solutions
presented in this paper are the relatively low signal-to-noise ratio of
the pulsars and the low detection rate, particularly in the
observations of the early 1990s. For some of the pulsars, the 20-cm
flux densities are so low that they are detectable only in about 10\%
of all observations (see Camilo et al.~2000\nocite{clf+00}).
Additionally, the data obtained between July 1989 and May 1991 were not
used in any of the fits because the difference between the Parkes clock
and UTC is not known to better than 50\,$\mu$s. This introduces
significant errors that would compromise several of the measurements.

\section{Pulsar positions}
\label{sec:positions}

\begin{figure}
\setlength{\unitlength}{1in}
\begin{picture}(0,3.5)
\put(0,0){\includegraphics{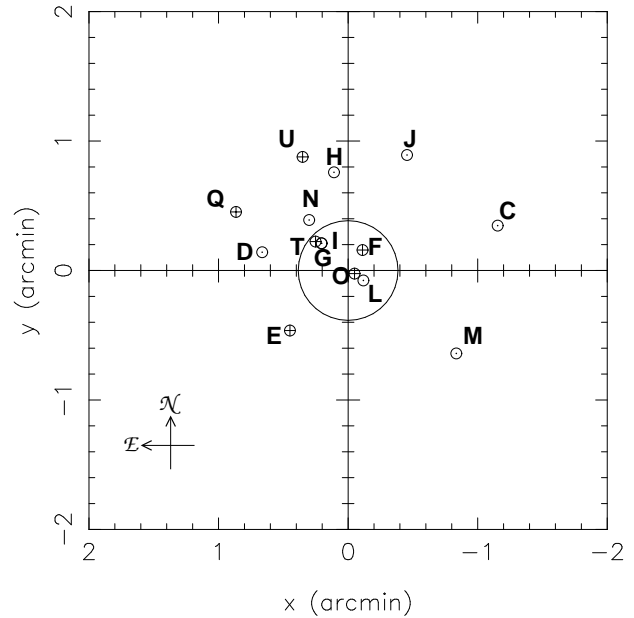}}
\end{picture}
\caption [] {Positions of the 15 pulsars in 47~Tuc with a timing
solution plotted in the plane of the sky.  The positions are given as
east-west (x) and north-south (y) offsets from the centre of the
cluster (see Table~\ref{tab:offsets}).  The pulsars indicated with
$\oplus$ have positive observed period derivatives ($\dot{P}$), while
those indicated with $\odot$ have negative $\dot{P}$.  The central
circle indicates the core radius. }

\label{fig:positions}
\end{figure}

The timing solutions presented in the previous section bring the number
of pulsars in 47~Tuc with accurately known positions to 15, and with
measured proper motions to five. For easy reference, we derive from the
coordinates in Table~\ref{tab:common_parameters} the east-west (x) and
north-south (y) offsets of the pulsars relative to the cluster centre.
These are presented in Table~\ref{tab:offsets} and depicted in
Fig.~\ref{fig:positions}. In the remainder of this section, we discuss
the implications of these positional measurements.

\begin{table}
\begin{center}
\begin{small}
\caption{East-west (x) and north-south (y) offsets of 15 pulsars from
the centre of the cluster, assumed here to be {\em exactly\/} at
$\alpha = \rm 00^{\rm h} 24^{\rm m} 05\fs29$ and $\delta = \rm
-72^\circ04'52\farcs3$. The errors in the offsets are 2 or less in the
last digits quoted.  The angular distance of each pulsar from the
centre of the cluster, $\theta_{\perp}$, has an accuracy limited by the
uncertain absolute position of the cluster
(Table~\protect\ref{tab:47tuc}). }
\begin{tabular}{ c l l c }
\multicolumn{4}{c}{}\\
\hline
Pulsar &
\multicolumn{1}{c}{x}             &
\multicolumn{1}{c}{y}             &
\multicolumn{1}{c}{$\theta_{\perp}$} \\
 & 
\multicolumn{1}{c}{(arcmin)}      &
\multicolumn{1}{c}{(arcmin)}      &
\multicolumn{1}{c}{(arcmin)}      \\
\hline
 C &  $-$1.1541 &   @0.3469   &    1.21 \\
 D &  @0.6634   &   @0.1411   &    0.68 \\
 E &  @0.4489   &   $-$0.4639 &    0.65 \\  
 F &  $-$0.1111 &   @0.1583   &    0.19 \\  
 G &  @0.2060   &   @0.2103   &    0.29 \\ 
 \multicolumn{4}{c}{}\\
 H &  @0.1088   &   @0.7585   &    0.77 \\  
 I &  @0.2041   &   @0.2106   &    0.29 \\  
 J &  $-$0.4547 &   @0.8921   &    1.00 \\  
 L &  $-$0.1174 &   $-$0.0767 &    0.14 \\  
 M &  $-$0.8347 &   $-$0.6404 &    1.05 \\  
\multicolumn{4}{c}{}\\
 N &  @0.3008   &   @0.3904   &    0.49 \\  
 O &  $-$0.0495 &   $-$0.0242 &    0.06 \\  
 Q &  @0.8651   &   @0.4525   &    0.98 \\  
 T &  @0.2512   &   @0.2232   &    0.34 \\  
 U &  @0.3509   &   @0.8772   &    0.94 \\  
\hline
\end{tabular}

\label{tab:offsets}
\end{small}
\end{center}
\end{table}

\subsection{Radial distribution}
\label{sec:radial}

The radial distribution for the pulsars in 47~Tuc was first addressed
by Rasio (2000)\nocite{ras00}, based on an earlier version of some of
the results we present here \cite{fcl+00}.  We now outline the main
characteristics of this distribution and draw some conclusions from
it.

The most striking characteristic of the pulsar distribution is that all
the pulsars in 47~Tuc with measured positions lie within $1\farcm2$ (3
core radii) of the centre of the cluster, despite the fact that the
tidal radius of 47~Tuc is about $40'$ (Table~\ref{tab:47tuc}).  This
distribution, as we shall now see, is not an artifact introduced by the
size and shape of the Parkes telescope radio beam pattern, which has a
half-power radius of $\sim 7'$ at 1400\,MHz.

The flux densities for most of the pulsars have been calculated by
Camilo et al. (2000). Among the pulsars with a known solution, 47~Tuc~N
has the lowest flux density (0.03 $\pm$ 0.01 mJy).  Supposing that this
value is the lower limit for the flux density of a pulsar for which we
can obtain a timing solution, weak pulsars like 47~Tuc~U, with a flux
density of 0.06 $\pm$ 0.01 mJy, are detectable in a circle with a
radius of at least $7'$.  There are 10 pulsars with at least this flux
density, and the fact that none of these is seen outside a radius of
$1\farcm2$ is therefore a true feature of the pulsar distribution, and
not an artifact due to the shape of the beam.  It should also be noted
that the original surveys at 660\,MHz \cite{mlr+91}, with a larger
telescope beam, covered an area approximately 100 times the size of the
roughly arcmin-scale central region in which we now localize the 15
pulsars.

\begin{figure}
\setlength{\unitlength}{1in}
\begin{picture}(0,3.2)
\put(0,0){\includegraphics{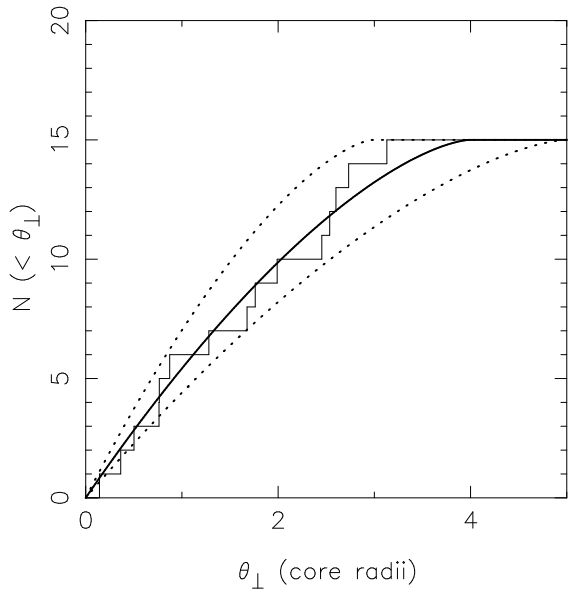}}
\end{picture}
\caption [] {Cumulative radial distribution of the pulsars in
47~Tuc. This is compared with the expectation of a distribution of the
type $n(r) \propto r^{-2}$ with a cutoff at 4 core radii (solid line)
and 3 and 5 core radii (dotted lines). }
\label{fig:radial}
\end{figure}

\begin{figure}
\setlength{\unitlength}{1in}
\begin{picture}(0,3.2)
\put(0,0){\includegraphics{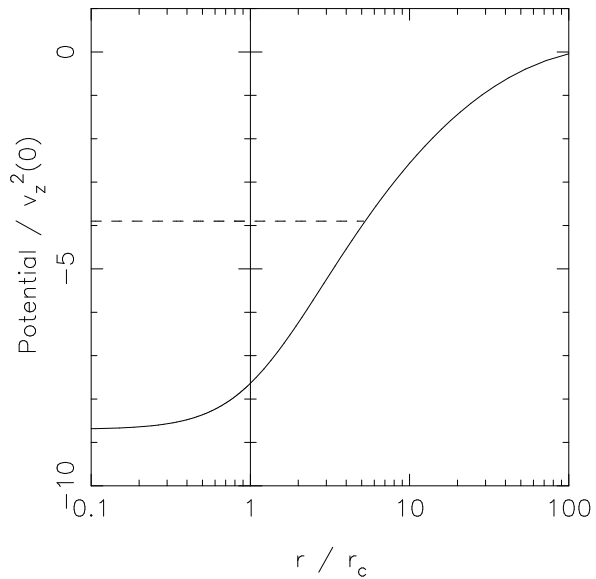}}
\end{picture}
\caption [] {Globular cluster potential as a function of distance from
the centre given by a King model of 47~Tuc. The maximum squared pulsar
velocity at the centre is indicated by the height of the dashed line
above $W(r=0) = -8.7 v_z^2(0)$. The pulsars cannot travel further than
about 5 core radii from the centre (see \S~\ref{sec:radial}). }
\label{fig:potential}
\end{figure}

The present pulsar distribution is essentially an equilibrium
distribution. The characteristic ages of these pulsars are of the order
of $10^9$ yr, while the relaxation time (the time it takes for a pulsar
to change kinetic energy significantly through interactions with other
stars) in the core of 47~Tuc is of the order of $10^8$ yr \cite{djo93}.
Therefore, the pulsars are expected to be in thermal equilibrium with
the remaining stars, and their distribution should be merely a function
of their mass and of the potential of the cluster.  Indeed, the
de-projected pulsar distribution is consistent with $n(r) \sim r^{-2}$
until about 4 core radii, decreasing sharply outside this radius (see
Fig.~\ref{fig:radial}).

This concentration of pulsars near the centre of the cluster is a
consequence of mass segregation: if all objects in a cluster have
reached thermal equilibrium, i.e., all stellar populations have a
similar average kinetic energy, then the most massive populations,
among which are the pulsars, will have smaller velocities, and
therefore will dwell deeper in the potential well of the cluster.  To
quantify this, we note that the kinetic energy of the lighter stars in
the cluster core, with mass $m_{ms}$, is limited by the escape velocity
$v_e \equiv \sqrt{- 2 W(0)}$.  Equating the resulting limit of kinetic
energy to that of the pulsars results in a limit to the velocity of
pulsars bound in the cluster core:
\begin{equation}
v_{p\,\rm max}^2 = (m_{ms} / m_{p}) v_e^2,
\label{eq:kinetic}
\end{equation}
where $m_p$ is the pulsar mass. As pulsars move away from the centre of
the cluster under the influence of the cluster's gravitational
potential their velocities decrease, eventually reaching zero at a
maximum possible distance $r_{\rm lim}$ from the centre of the
cluster.

The potential energy $U(r)$ at $r_{\rm lim}$ is calculated from
conservation of energy:
\begin{equation}
U(r_{\rm lim}) \equiv m_p W(r_{\rm lim}) = m_p W(0) + m_p v_{p\,\rm max}^2/2,
\end{equation}
where $W(r)$ is the cluster potential. Using equation~\ref{eq:kinetic}
we obtain
\begin{equation}
W(r_{\rm lim}) = W(0) \left( 1 - \frac{m_{ms}}{m_{p}}\right).
\end{equation}
This limit, with $m_{ms} = 0.8\,\rm M_{\odot}$ and $m_{p} = 1.45\,\rm
M_{\odot}$ (the average mass for isolated and binary pulsars), is
represented by a dashed line in Fig.~\ref{fig:potential}, where we use
a King model for the gravitational potential of 47~Tuc (King 1966;
Phinney, priv. comm.)\nocite{kin66}. From this we derive $r_{\rm lim}
\sim 5 r_c$, which is very close to the inferred 4 $r_c$. Therefore,
the confinement of the pulsars in the inner region can be explained, to
first order, as resulting from the shape of the potential and the upper
limit for the kinetic energy of the pulsars. The fact that all pulsars
are located close to the core is a constraint on any future more
accurate model of the gravitational potential of the cluster.

Spitzer~(1987)\nocite{spi87} demonstrates that the distributions of two
stellar species of mass $m_i$ and $m_j$ in thermal equilibrium with all
stars in a cluster are related by
\begin{equation}
n_i (r) \propto n_j(r)^{m_i/m_j}.
\label{eq:r_distribution}
\end{equation}
Therefore, two stellar populations with similar masses in thermal
equilibrium should have similar radial distributions.  Some of the most
massive blue stragglers known in 47~Tuc have masses of
1.3--1.6\,M$_{\odot}$ \cite{gbe+98}. Because this is the mass range
expected for pulsars \cite{tc99}, including binaries such as those in
47~Tuc \cite{clf+00}, both populations should have similar radial
distributions, which is apparently the case \cite{ras00}.  However we
caution that the population considered by Rasio (2000)\nocite{ras00} is
not altogether homogeneous, having a relatively wide range of masses.
Ultimately it will be useful to compare the radial distribution of just
the most massive blue stragglers with that of pulsars, and in
particular to determine whether they display the abrupt radial cutoff
observed for pulsars.

Lastly, Spitzer shows that outside 3 core radii, the mass distribution
for the dominant stellar species should be of the type $n(r) \propto
r^{-2}$. Knowing this and the fact that the radial distribution of the
pulsars in M15 is $n(r) \propto r^{-3.1}$ \cite{and92}, Phinney~(1992)
used equation~\ref{eq:r_distribution} to infer that the dominant
stellar species near the centre of M15 is 0.9~M$_{\odot}$ white dwarfs.
In principle we could use the same argument to show, based on the
observed pulsar distribution in 47~Tuc, that the dominant stellar
species in the central region of this cluster is made of
1.3--1.6\,M$_{\odot}$ objects. However, the fact that the pulsars lie
within (and not outside) 3 core radii may complicate this
interpretation (although the ``pulsar core'' should be more compact
than that of the cluster).

\begin{figure}
\setlength{\unitlength}{1in}
\begin{picture}(0,3.5)
\put(0,0){\includegraphics{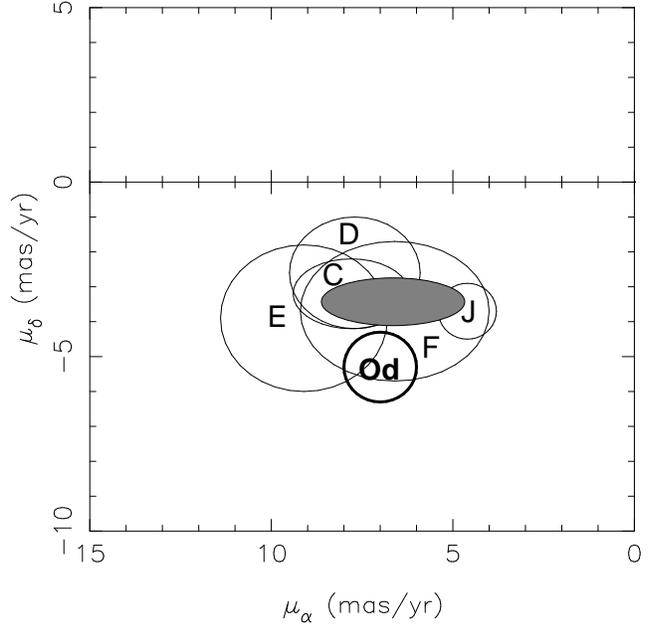}}
\end{picture}
\caption [] {Measured proper motions of five pulsars in 47~Tuc.  The
semi-axes of the ellipses represent the 1-$\sigma$ errors for the
proper motions quoted in Table~\ref{tab:proper_motions}. The value for
the proper motion of the cluster \cite{obgt97} is indicated by the
thick circle. The filled ellipse indicates the motion of the cluster
and its error deduced by averaging the measured pulsar proper motions.}
\label{fig:pm}
\end{figure}

\subsection{A triple system with two pulsars?}
\label{sec:triple}

There is a remarkable coincidence between the celestial coordinates of
47~Tuc~G, an isolated pulsar, and 47~Tuc~I, a binary pulsar (see
Table~\ref{tab:offsets}). The projected angular distance between these
two pulsars is only 120\,mas.

To estimate the chance probability of finding two pulsars in such close
proximity, suppose that pulsars are randomly distributed in a disk of
radius $18''$ (the projected angular distance of these two pulsars from
the centre of the cluster). For a given pulsar, the probability of
finding a second one at a projected distance of $0\farcs12$ or less is
proportional to the area of a disk of radius $0\farcs12$  divided by
the area of a disk of radius $18''$ or $\sim 5 \times 10^{-5}$.  The
probability of being at a distance larger than $0\farcs12$ is 0.99995.
For the five pulsars known within $18''$, we have 10 different possible
pairs. The probability of not finding any pulsar within $0\farcs12$ of
any other pulsar is $0.99995^{10} = 0.9995$, i.e., the overall
probability of finding one or more pulsars projected within $0\farcs12$
of any other is $5 \times 10^{-4}$, about one in 2000.

If the pulsars are near to each other but not interacting
significantly, then the acceleration caused by the cluster should be
similar for both of them.  In \S~\ref{sec:accel} we see that their
values of $\dot{P}/P$ are similar, with a difference of
$3\times10^{-18}$\,s$^{-1}$, in a possible range $\sim$ 20 times
larger.  Therefore, the combined probability of finding the pulsars
with such close projected separation and with such similar
accelerations is about one in 40,000.

Given this low formal probability of chance coincidence, it is worth
considering that these two pulsars may be in a hierarchical triple
system with a major axis of at least 600\,a.u.  A system with
two components 600\,a.u. apart is not likely to remain bound for long
in the dense environment of a globular cluster. By considering the
stellar flux with one-dimensional speeds of $v_z(r)$, an estimate for
the mean time a binary system can survive in a dense environment (i.e.,
before it gets hit by another cluster member) is
\begin{equation}
\tau = (n[r] \sigma \sqrt{3} v_{z}[r])^{-1}.  
\label{eq:lifetime}
\end{equation}
In this expression $n(r)$ is the density of stars, $v_z(r)$ is the
one-dimensional stellar velocity dispersion near the binary and
$\sigma$ is the cross-sectional area for the interaction of the binary
with a star from the cluster.  The King model for 47~Tuc predicts $n
(r) \sim 4 \times 10^4$\,pc$^{-3}$ for this location.  With $v_z(r)
\sim 10$\,km\,s$^{-1}$ we estimate $\tau \sim 10^4$--$10^5$\,yr.  This
time is comparable to the orbital period of the putative binary system,
and $10^{-6}$--$10^{-5}$ the age of the cluster.  Thus, despite the low
formal probability of a chance coincidence, this simple calculation
argues against the reality of such an association.  Ultimately, the
reality or otherwise of this system can be tested by future
measurements of higher period-derivatives for the two pulsars.

\subsection{Proper motions}
\label{sec:pm}

The proper motions listed in Table~\ref{tab:proper_motions} are
displayed in Fig.~\ref{fig:pm}, together with the value measured for
the proper motion of the cluster by Odenkirchen et al.
(1997)\nocite{obgt97}. At the 2-$\sigma$ level all these proper motions
are consistent with each other.

If a pulsar is moving with a transverse velocity $v_{p}$ (km\,s$^{-1}$)
relative to the centre of the cluster, the difference in proper motions
is $\Delta \mu=0.21 v_p/D$\,mas\,yr$^{-1}$, where $D$ is the distance
to the cluster in kpc. The escape velocity of 47~Tuc is about
58\,km\,s$^{-1}$ (Table~\ref{tab:47tuc}).  Therefore, in order to
detect a pulsar in an escape trajectory at the 3-$\sigma$ level, we
would have to measure relative proper motions of about
2.4\,mas\,yr$^{-1}$ with a precision better than 0.8\,mas\,yr$^{-1}$,
and we are approaching this level for 47~Tuc~J
(Table~\ref{tab:proper_motions}).  If the pulsars are in thermal
equilibrium with the surrounding stars, their relative motions are of
the order of the central velocity dispersion or less, i.e., $\sim
12$\,km\,s$^{-1}$.  Thus, detecting their relative motions implies
measuring relative proper motions of about 0.5\,mas\,yr$^{-1}$ with a
precision of 0.15\,mas\,yr$^{-1}$. The typical precision for the
positions of the weakest pulsars is around 5\,mas. In order to obtain a
precision of 0.15\,mas\,yr$^{-1}$ for these we would have to measure
the positions of the pulsars again in $\sim 50$ years' time. However,
for the brightest pulsars, with positional accuracies of about 1\,mas,
it will suffice to make another measurement in about 10 years, using
current data-acquisition systems.

The assumption of thermal equilibrium for the pulsars implies that,
with current timing precision, a measurement of the proper motions of
these pulsars is in effect a measurement of the proper motion of the
cluster.  We can therefore make a weighted average of the five proper
motions and determine the motion of the cluster, knowing that the
peculiar motions of the pulsars should be smaller than the errors in
the individual measurements.  The weights chosen are simply the inverse
of the uncertainties in each coordinate.  For the currently-observed
proper motions, we find the weighted average value to be $\mu_{\alpha}
= (6.6 \pm 1.9)$\,mas\,yr$^{-1}$ and $\mu_{\delta} = (-3.4 \pm
0.6)$\,mas\,yr$^{-1}$.  The uncertainties are taken to be the sum in
quadrature of the weighted dispersion of the values of proper motion
for the pulsars about the average, and 0.5\,mas\,yr$^{-1}$, which
accounts for the expected actual dispersion of proper motions.

\section{Pulsar accelerations in the cluster potential}
\label{sec:accel}

\begin{figure}
\setlength{\unitlength}{1in}
\begin{picture}(0,3.2)
\put(0.1,0){\includegraphics{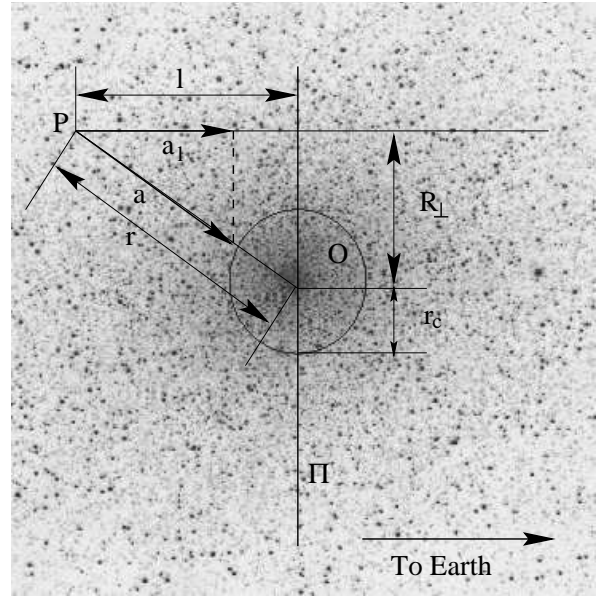}}
\end{picture}
\caption [] {Geometrical parameters used in \S~\ref{sec:accel}:  $\Pi$
is a plane perpendicular to the line of sight that passes through the
centre of the cluster, at point O, with core radius $r_{c}$.  For a
pulsar at point P, $R_{\perp}$ is the projected distance to, and $a$ is
the acceleration of the pulsar towards, the centre of the cluster.  The
line-of-sight component of $a$, $a_{l}$, is the only one detectable
from the Earth (located in the plane of the figure). }

\label{fig:cluster}
\end{figure}

Nine of the 15 pulsars in Table~\ref{tab:common_parameters} have
negative period derivatives ($\dot{P}$). This has been observed before
for some pulsars located in globular clusters (e.g., Wolszczan et al.
1989)\nocite{wkm+89}. Rather than being due to intrinsic spin-up,
negative period derivatives are thought to be caused by the
acceleration of the pulsar towards the Earth in the cluster potential
(see Fig.~\ref{fig:cluster}).

In \S~\ref{sec:rhovz0} we use the observed period derivatives of the
pulsars to estimate lower limits for the surface mass density of the
cluster.  We then consider the aptness of a King model together with
accepted parameters for 47~Tuc in describing the accelerations
experienced by the pulsars as inferred from the period derivatives
(\S~\ref{sec:king}).  We derive limits for the ages and magnetic fields
of the pulsars in \S~\ref{sec:47tucpsrlimits}.

The observed period derivative, $\dot P_{\rm obs}$, is the sum of the
pulsar's intrinsic spin-down, $\dot P_{\rm int}$, and the effect of the
acceleration along the line of sight, $a_l$.  This sum may be negative
if a negative $a_l$ contribution is not exceeded by a positive $\dot
P_{\rm int}$; i.e., a negative $\dot P_{\rm obs}$ implies $|a_l/c| >
\dot P/P_{\rm int}$.  It also implies $|a_l/c| > |\dot P/P_{\rm
obs}|$.

The acceleration along the line of sight has several components,
denoted by $a_S$, $a_G$, and $a_C$ in
\begin{equation}
\left( \frac{\dot{P}}{P} \right)_{\rm obs} = 
\frac{a_{S}}{c} +
\frac{a_{G}}{c} +
\frac{a_{C}}{c} + 
\left( \frac{\dot{P}}{P} \right)_{\rm int},
\label{eq:formula1}
\end{equation}
where the terms represent:

1. The centrifugal acceleration \cite{shk70}, given by
\begin{equation}
\frac{a_{S}}{c} = \frac{\mu^{2}D}{c}.
\label{eq:Shkl}
\end{equation}
This term, using the values in Table~\ref{tab:47tuc}, amounts to
$a_{S}/c = (+9 \pm 3) \times 10^{-19}$\,s$^{-1}$.

2. The difference in Galactic acceleration along the line of sight
between a given object and the barycentre of the solar system
($a_{G}$). This is a function of the object's Galactic latitude $b$,
longitude $l$, and distance from the solar system $D$. For 47~Tuc we
have $b = -44.9^\circ$, $l = 305.9^\circ$ and $D = 5.0 \pm 0.4$\,kpc.
Using Paczynski's~(1990)\nocite{pac90} model of the gravitational
potential of the Galaxy, we obtain $a_{G}/c = (-4.5 \pm 0.2) \times
10^{-19}$\,s$^{-1}$.

3. Accelerations due to the gravitational field of the globular cluster
and its individual stars ($a_{C}$). This is the most interesting
contribution from an astrophysical point of view.

Contributions 1 and 2 total $(+5 \pm 3) \times 10^{-19}$\,s$^{-1}$.
Henceforth, $(\dot{P}/{P})_{\rm obs}$ indicates the measured value of
$\dot{P}/{P}$ minus these contributions, and $a_l$ refers solely to
$a_C$. All conclusions in this section apply only to the pulsars with
known timing solutions.

\subsection{Projected surface mass density of 47 Tuc}
\label{sec:rhovz0}

In Fig.~\ref{fig:cluster} are shown some useful geometrical parameters
we use in the remainder of this section.  The plane $\Pi$ passes
through the centre of the cluster and is perpendicular to the line of
sight, and the core radius is given by $r_c = D \theta_c $.  A
particular observed pulsar has unknown distance to the centre of the
cluster, $r$, and to $\Pi$, $l$ --- we know only the projected distance
of the pulsar to the centre of the cluster, $R_{\perp} = D
\theta_{\perp}$.  The acceleration along the line of sight, $a_l$, is
the only component potentially detectable from the Earth.

We now derive constraints for the surface mass density of the cluster
in its inner regions.  According to Phinney (1993)\nocite{phi93}, the
following expression is valid to within $\sim 10$\% for all plausible
cluster models, and for all $\theta_{\perp}$:
\begin{eqnarray}
\frac{a_{l\,{\rm max}}(\theta_{\perp})}{c} \simeq
1.1 \frac{G M_{\rm cyl} (<\theta_{\perp})}{c \pi D^2 \theta_{\perp}^2} = \nonumber \\
5.1\times10^{-19} \left( 
\frac{ \overline{\Sigma} (<\theta_{\perp}) }{10^3\,\rm M_{\odot}\,pc^{-2} }
 \right)\,\rm s^{-1},
\label{eq:sdensity}
\end{eqnarray}
where $M_{\rm cyl} (<\theta_{\perp})$ is the mass of all the matter
with a projected distance smaller than $\theta_{\perp}$, and
$\overline{\Sigma} (<\theta_{\perp})$ is the corresponding projected
surface mass density.  The limit on $\overline{\Sigma}$ does not depend
on estimates of the distance to the cluster or core radius. The pulsars
with negative observed period derivatives provide a lower bound on
$a_{l\,{\rm max}}(\theta_{\perp})$ and therefore, with
equation~\ref{eq:sdensity}, on $\overline{\Sigma}$ and $M_{\rm cyl}$.
These limits are presented in Table~\ref{tab:params47tuc}.

\begin{table}
\begin{center}
\caption{Constraints on projected surface mass density for 47~Tuc
obtained from nine pulsars with negative $(\dot{P}/P)_{\rm obs}$ (see
\S~\ref{sec:rhovz0}). The distance to the cluster is taken to be
5.0\,kpc. }
\begin{tabular}{ c l l l l}
\hline
Pulsar &
\multicolumn{1}{c}{$(\dot{P}/P)_{\rm obs}$}          &
\multicolumn{1}{c}{$R_{\perp}$}			     &
\multicolumn{1}{c}{$\overline{\Sigma} (<\theta_{\perp})$} &
\multicolumn{1}{c}{$M_{\rm cyl} (<\theta_{\perp})$}	     \\
    &
\multicolumn{1}{c}{($10^{-17}\,\rm s^{-1}$)} 	             &
\multicolumn{1}{c}{($\rm pc$)}			             &
\multicolumn{1}{c}{($10^{3}\,{\rm M}_{\odot}\,\rm pc^{-2}$)} &
\multicolumn{1}{c}{($10^{3}\,{\rm M}_{\odot}$)}	             \\
\hline
  C & $-$0.91 & 1.75  & $>18$ & $>170$    \\
  D & $-$0.11 & 0.99  & $>2.2$  & $>6.6$  \\
  G & $-$1.09 & 0.43  & $>21$ & $>12$     \\
  H & $-$0.10 & 1.11  & $>1.9$& $>7.5$    \\
  I & $-$1.36 & 0.43  & $>27$ & $>15$     \\
\multicolumn{5}{c}{}\\
  J & $-$0.51 & 1.46  & $>10$& $>67$     \\
  L & $-$2.85 & 0.20  & $>56$ & $>7.3$    \\
  M & $-$1.09 & 1.53  & $>21$ & $>160$    \\
  N & $-$0.76 & 0.72  & $>15$ & $>24$     \\
\hline
\end{tabular}

\label{tab:params47tuc}
\end{center}
\end{table}

\subsection{Accounting for the pulsar accelerations}
\label{sec:king}

\begin{table*}
\begin{center}
\caption{Comparison of estimates of ``observed'' pulsar accelerations
$A_o$ with predicted average accelerations, $A_a(R_{\perp})$, derived
using a King model for 47~Tuc with $D = 5.0\pm0.4$\,kpc, $v_z(0) =
11.6\pm1.4$\,km\,s$^{-1}$ and $\theta_c = 23\farcs1$ (see
\S~\ref{sec:king}).  The maximum acceleration predicted by the model
for each projected pulsar offset from the centre of the cluster, used
in Fig.~\ref{fig:c1}, is also listed. }
\begin{tabular}{ c c c c c }
\hline
Pulsar	                                            &
\multicolumn{1}{c}{$|A_o/c|$}                       &
\multicolumn{1}{c}{$|A_a(R_{\perp})/c|$}            &
\multicolumn{1}{c}{$|a_{l\,\rm max}(R_{\perp})/c|$} &
\multicolumn{1}{c}{$|A_o/A_a(R_{\perp})|$}          \\
                                                    &
\multicolumn{1}{c}{($10^{-17}\,\rm s^{-1}$)}        &
\multicolumn{1}{c}{($10^{-17}\,\rm s^{-1}$)}        &
\multicolumn{1}{c}{($10^{-17}\,\rm s^{-1}$)}        & 
					            \\
\hline
 C & 1.38 & $0.6\pm0.2$ & $1.0\pm0.3$ & $2.4\pm1.0$ \\
 D & 0.57 & $1.3\pm0.5$ & $1.8\pm0.7$ & $0.4\pm0.2$ \\
 E & 2.27 & $1.3\pm0.5$ & $1.9\pm0.7$ & $1.7\pm0.7$ \\
 F & 1.95 & $2.1\pm0.8$ & $3.6\pm1.3$ & $0.9\pm0.4$ \\
 G & 1.56 & $2.0\pm0.7$ & $3.2\pm1.1$ & $0.8\pm0.3$ \\
\multicolumn{5}{c}{}\\			        
 H & 0.56 & $1.1\pm0.4$ & $1.6\pm0.6$ & $0.5\pm0.2$ \\
 I & 1.83 & $2.0\pm0.7$ & $3.2\pm1.1$ & $0.9\pm0.4$ \\
 J & 0.98 & $0.8\pm0.3$ & $1.2\pm0.5$ & $1.2\pm0.5$ \\
 L & 3.32 & $2.0\pm0.7$ & $3.8\pm1.4$ & $1.6\pm0.7$ \\
 M & 1.55 & $0.8\pm0.3$ & $1.2\pm0.4$ & $2.0\pm0.9$ \\
\multicolumn{5}{c}{}\\			        
 N & 1.23 & $1.6\pm0.6$ & $2.4\pm0.8$ & $0.8\pm0.3$ \\
 O & 0.63 & $1.4\pm0.5$ & $4.0\pm1.4$ & $0.5\pm0.2$ \\
 Q & 0.33 & $0.9\pm0.3$ & $1.3\pm0.4$ & $0.4\pm0.2$ \\
 T & 3.37 & $2.0\pm0.7$ & $3.0\pm1.0$ & $1.7\pm0.7$ \\
 U & 1.68 & $0.9\pm0.3$ & $1.3\pm0.5$ & $1.9\pm0.7$ \\
\hline
\end{tabular}

\label{tab:accelerations}
\end{center}
\end{table*}

We have used a King model of 47~Tuc to calculate the gravitational
potential as a function of distance to the centre of the cluster,
$W(r/r_c)$, divided by $v_z(0)^2$ (Fig.~\ref{fig:potential}).  The
input for such a model is a single parameter, the logarithm of the core
radius divided by the tidal radius \cite{kin66}.

The radial derivative of $W(r/r_c)/v_z(0)^2$ yields a ``normalized
acceleration'', $a_n(r/r_c)$. In order to calculate accelerations $A$
in m s$^{-2}$, we need two more cluster parameters: the central
dispersion of line-of-sight velocities, $v_z(0)$, and the core radius,
$r_c \equiv D \theta_c$.  $A$ is obtained from
\begin{equation}
A(r/r_c) = \frac{a_{n}(r/r_c) v_{z}^2(0)}{D \theta_c}.
\label{eq:nacc}
\end{equation}

To calculate the contribution to $(\dot{P}/P)_{\rm obs}$ of the
acceleration of a pulsar at a given radius we multiply $A(r/r_c)$ by
$l/r$, so as to obtain the line-of-sight component of the
acceleration.  We do not, of course, know the actual distance $r$ of
each pulsar from the cluster centre, only its projection $R_\perp$.  We
will therefore compare the ``observed acceleration'' of each pulsar to
the ensemble average along the relevant lines of sight calculated from
the King model, taking the pulsar distribution in the cluster to have a
density proportional to $r^{-2}$ and a cut-off radius of 4 core radii
(see \S~\ref{sec:radial}).

The average acceleration along each line of sight (integrated over only
one hemisphere) is thus given by
\begin{equation}
A_a (R_{\perp}) = \frac{1}{N}\int_{l = 0}^{ l = \sqrt{r_{\rm lim}^2 -
R_{\perp}^2} } \frac{\lambda^{-1}}{r^2} A(r/r_c) \frac{l}{r} dl,
\label{eq:AvAcc}
\end{equation}
where $\lambda$ is a normalizing constant such that $\lambda^{-1}/r^2$
is the local linear density of simulated pulsars, and $N$ is their
total number in a column along the integration path.

How do these ``average'' theoretical accelerations along each line of
sight compare with the observed accelerations?  The observed
acceleration, $A_o$, can be obtained for each pulsar by subtracting the
intrinsic $\dot{P}/P$ from $(\dot{P}/P)_{\rm obs}$.  Unfortunately, the
intrinsic period derivative is not known.   We derive limits on this
quantity in \S~\ref{sec:47tucpsrlimits}, where we also find that the
average characteristic age of the pulsars is greater than $\sim
1$\,Gyr.  Here we assume that each pulsar has a characteristic age of
3\,Gyr and subtract the corresponding $\dot{P}/P = +0.5 \times
10^{-17}$\,s$^{-1}$ from each observed value (listed in
Table~\ref{tab:47tucpsrlimits}) to obtain an estimate for $A_o$. These,
along with values of $A_a$ calculated from equation~\ref{eq:AvAcc}, are
presented in Table~\ref{tab:accelerations}.

We are now in a position to compare the ``observed'' and average
predicted accelerations.  We do this for each pulsar in the last column
in Table~\ref{tab:accelerations}.  This value is significantly biased
for some individual pulsars by the subtraction of a fixed $\dot{P}/P =
0.5 \times 10^{-17}$\,s$^{-1}$, our first-order attempt at accounting
for the intrinsic period derivatives.  However, the average of all
values in the last column of the table should be approximately unity,
if the modeling described above is to be consistent with the actual
accelerations experienced by the pulsars --- which indeed it is,
despite considerable uncertainty:  $\overline{|A_o/A_a(R_{\perp})|} =
1.2 \pm 0.7$.

\begin{figure}
\setlength{\unitlength}{1in}
\begin{picture}(0,3.2)
\put(0,0){\includegraphics{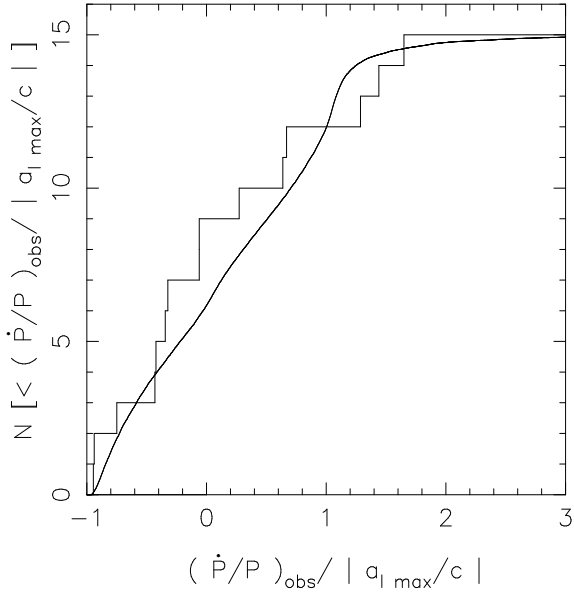}}
\end{picture}
\caption [] {Cumulative distributions of $(\dot{P}/P)_{\rm
obs}/|a_{l\,\rm max}/c|$ for 15 observed pulsars, and for a simulated
population (smooth curve), normalized by the observed distribution (see
\S~\ref{sec:king}). }

\label{fig:c1}
\end{figure}

As an additional consistency check on the King model of the cluster
potential we performed a Monte Carlo simulation of the pulsar
population of 47~Tuc. The aim of this simulation is to see, given
reasonable input assumptions, whether we can reproduce the {\em
distribution\/} of $(\dot{P}/P)_{\rm obs}$.

In the simulation, we generate a spherically symmetric population of
$10^6$ pulsars with a radial distribution of the type $n(r) \propto
r^{-2}$ and a cutoff of 4 core radii.  Each of the pulsars was randomly
assigned an age from a flat distribution ranging between 500\,Myr and
10\,Gyr --- the mean $\dot{P}/P = 0.5 \times 10^{-17}$\,s$^{-1}$
corresponding to what was assumed before.  These assigned ages
correspond to the intrinsic $\dot{P}/P$ of each of the model pulsars.
Finally, given the position of each of the model pulsars relative to
the cluster centre, we calculate the contribution to $\dot{P}/P$ from
the acceleration in the cluster determined from the King model with $D
= 5.0$\,kpc, $v_z(0) = 11.6$\,km\,s$^{-1}$ and
$\theta_c\,=\,23\farcs1$.
 By adding this contribution to the intrinsic $\dot{P}/P$ we have, for
each pulsar, a model observed $\dot{P}/P$ which can then be directly
compared to the sample of 15 pulsars for which we have timing
solutions.

The results of this simulation are shown in Fig.~\ref{fig:c1}, where we
present a cumulative plot of $(\dot{P}/P)_{\rm obs}/|a_{l\,\rm
max}(R_{\perp})/c|$ for the real and simulated samples.  Given the
relatively small size of the observed distribution, and the
straightforward simulation that we have performed, the agreement
between the model and observed samples is good.

Our overall conclusion is therefore that a King model for 47~Tuc, with
$D = 5.0 \pm 0.4$\,kpc, $v_z(0) = 11.6 \pm 1.4$\,km\,s$^{-1}$,
$\theta_c\,=\,23\farcs1$ and a small contribution from the intrinsic
$\dot{P}/P$ of the pulsars, provides a good description for the
observed values of $\dot{P}/P$.

\begin{figure}
\setlength{\unitlength}{1in}
\begin{picture}(0,3.5)
\put(0,0){\includegraphics{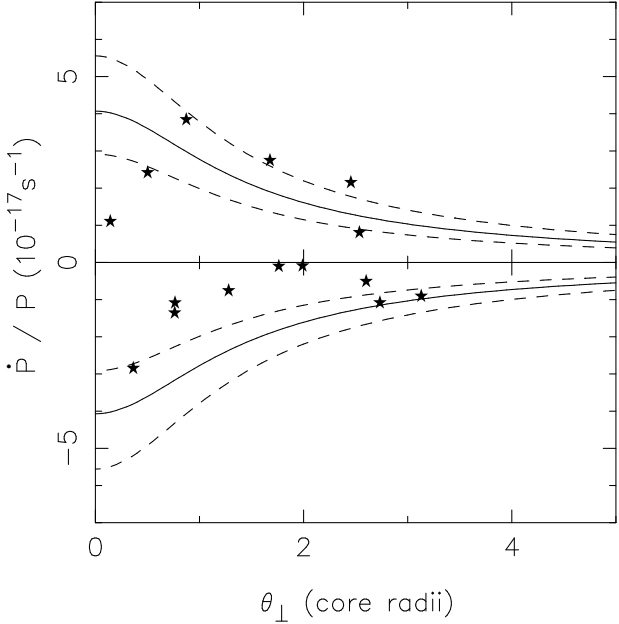}}
\end{picture}
\caption [] {The values for $(\dot{P}/P)_{\rm obs}$
(Table~\ref{tab:47tucpsrlimits}) are plotted versus $\theta_{\perp}$
for the 15 pulsars with coherent timing solutions (stars).  We also
plot the maximum acceleration along each line of sight calculated from
a King model with the nominal distance and central velocity dispersion
(solid line), and accounting for the uncertainties in those quantities
(dashed lines). For all the curves and pulsar positions, we assumed
$\theta_c$ to have its nominal value. See Table~\ref{tab:47tuc} for
cluster parameters and their uncertainties.}

\label{fig:accel}
\end{figure}

\subsection{Limits on pulsar ages and magnetic fields}
\label{sec:47tucpsrlimits}

We now obtain one-sided limits for the characteristic age and the
inferred surface dipole magnetic field strength of the pulsars, by
calculating the maximum intrinsic period derivative for each pulsar,
$\dot P_{\rm int\,max}$.  First, suppose we have a reliable estimate
for the maximum possible acceleration for each pulsar's line of sight,
$a_{l\,\rm max}(R_{\perp})$.  Next, assume that each pulsar actually
experiences such a maximum (negative) acceleration.  Under such
conditions, an observed $|\dot{P}/P|$ that is different from
$|a_{l\,\rm max}(R_{\perp})/c|$ is caused by a finite (positive)
intrinsic $\dot{P}/P$.  Hence, $(\dot{P}/P)_{\rm int} < |a_{l\,\rm
max}(R_{\perp})/c| + (\dot{P}/P)_{\rm obs}$.

After determining $\dot P_{\rm int\,max}$ we derive a minimum
characteristic age of the pulsar, $\tau_{\rm c} > P/(2 \dot P_{\rm
int\,max})$, and an upper limit for the magnetic field, $B < 3.2 \times
10^{19} (P \dot P_{\rm int\,max})^{1/2}$\,Gauss.

Only a reliable upper limit for the acceleration along each line of
sight remains to be calculated.  Phinney~(1993) calculated the maximum
acceleration expected near the centre of a globular cluster,
\begin{equation}
a_{l\,{\rm max}}(R_{\perp}) = \frac{3}{2} \frac{v_{z}^{2}(R_{\perp})} 
{\sqrt{ r_{c}^{2}+R_{\perp}^{2}}}.
\label{eq:almax2}
\end{equation}
This expression is valid to within $\sim 10$\% for $R_{\perp} \la 2
r_c$, provided that the function $v_{z}(R_{\perp})$ is accurately
known.

The values of $\dot P/P$ observed for the pulsars (reflecting in part
the gravitational potential of the cluster) do not appear to decrease
greatly with $R_{\perp}$ (Fig.~\ref{fig:accel}), so we first assume
conservatively that $a_{l\,\rm max}(R_{\perp})$ does not decrease with
$R_{\perp}$.  Therefore, the absolute upper limit for $a_{l\,\rm
max}(R_{\perp})$ is $a_{l\,\rm max}(0)$.  Within the constraints
imposed by the parameters and uncertainties in Table~\ref{tab:47tuc},
the maximum possible value for $a_{l\,\rm max}$ at the centre of the
cluster is obtained with $D=4.6$\,kpc, $v_z(0)=13.0$\,km\,s$^{-1}$ and
$\theta_c=21\farcs4$: according to equation~\ref{eq:almax2}, $a_{l\,\rm
max}/c=5.74\times10^{-17}$\,s$^{-1}$.

\begin{table*}
\begin{center}
\caption{Limits on period derivative, characteristic age and magnetic
field for the pulsars in 47~Tuc.  The $(\dot P/P)_{\rm obs}$ have been
corrected for the acceleration terms $a_{S}$ and $a_{G}$, using
$D=4.6$\,kpc. The values annotated with ``2'' were derived from the
maximum accelerations (also indicated) calculated with an isotropic
single-mass King model for 47~Tuc with the same value for the cluster
distance, $v_z(0)=13.0$\,km\,s$^{-1}$ and the nominal value for core
radius (see \S~\ref{sec:47tucpsrlimits}).  The values annotated with
``1'' were calculated with the same parameters, except for angular core
radius, where the lower limit, $\theta_c=21\farcs4$, was used.  In this
case $a_{l\,\rm max}(R_{\perp})/c = a_{l\,\rm max}(0)/c = 5.74 \times
10^{-17}$\,s$^{-1}$ (eq.~\ref{eq:almax2}).}
\begin{tabular}{ c c c c c c c l l l l l l }
\hline
Pulsar	&
\multicolumn{1}{c}{$P$}	&
\multicolumn{1}{c}{$(\dot{P}/P)_{\rm obs}$}	&
\multicolumn{1}{c}{$\dot{P}_{\rm int\,1}$}	&
\multicolumn{1}{c}{$\tau_{\rm c\,1}$}		&
\multicolumn{1}{c}{$B_1$}			&
\multicolumn{1}{c}{$|a_{l\,\rm max}/c|$}	&
\multicolumn{2}{c}{$\dot{P}_{\rm int\,2}$}	&
\multicolumn{2}{c}{$\tau_{\rm c\,2}$}		&
\multicolumn{2}{c}{$B_2$}			\\
&
\multicolumn{1}{c}{(ms)} 	&
\multicolumn{1}{c}{($10^{-17}\,\rm s^{-1}$)} 	&
\multicolumn{1}{c}{($10^{-19}$)}	&
\multicolumn{1}{c}{($10^{8}$\,yr)} 		&
\multicolumn{1}{c}{($10^{9}$\,G)}		&
\multicolumn{1}{c}{($10^{-17}\,\rm s^{-1}$)} 	&
\multicolumn{2}{c}{($10^{-19}$)} 		&
\multicolumn{2}{c}{($10^{8}$\,yr)} 		&
\multicolumn{2}{c}{($10^{9}$\,G)}		\\
\hline
C & 5.757 & $-$0.91 & $<2.8$ & $>3.3$ & $<1.3$ & 1.33 & $<0.24$ &\dots & $>38$  & \dots & $<0.4$ & \dots  \\
D & 5.358 & $-$0.10 & $<3.0$ & $>2.8$ & $<1.3$ & 2.48 & $<1.3 $ &\dots & $>6.7$ & \dots & $<0.8$ & \dots  \\
E & 3.536 & @2.74   & $<3.0$ & $>1.9$ & $<1.0$ & 2.60 & $<1.9 $ & $>0.05$& $>3.0$ & $<110$ & $<0.8$ & $>0.14$ \\
F & 2.624 & @2.42   & $<2.1$ & $>1.9$ & $<0.8$ & 4.92 & $<1.9 $ &\dots & $>2.2$ & \dots & $<0.7$ & \dots  \\
G & 4.040 & $-$1.09 & $<1.9$ & $>3.4$ & $<0.9$ & 4.32 & $<1.3 $ &\dots & $>4.9$ & \dots & $<0.7$ & \dots  \\
\multicolumn{9}{c}{}\\                	        	              
H & 3.210 & $-$0.09 & $<1.8$ & $>2.8$ & $<0.8$ & 2.21 & $<0.7 $ &\dots & $>7.5$ & \dots & $<0.5$ & \dots  \\
I & 3.485 & $-$1.36 & $<1.5$ & $>3.6$ & $<0.7$ & 4.32 & $<1.0 $ &\dots & $>5.3$ & \dots & $<0.6$ & \dots  \\
J & 2.101 & $-$0.51 & $<1.1$ & $>3.0$ & $<0.5$ & 1.66 & $<0.24$ &\dots & $>14$  & \dots & $<0.2$ & \dots  \\
L & 4.346 & $-$2.85 & $<1.3$ & $>5.5$ & $<0.8$ & 5.20 & $<1.0 $ &\dots & $>6.8$ & \dots & $<0.7$ & \dots  \\
M & 3.677 & $-$1.08 & $<1.7$ & $>3.4$ & $<0.8$ & 1.57 & $<0.18$ &\dots & $>32$  & \dots & $<0.3$ & \dots  \\
\multicolumn{9}{c}{}\\                	        	              
N & 3.054 & $-$0.76 & $<1.5$ & $>3.2$ & $<0.7$ & 3.23 & $<0.8 $ &\dots & $>6.4$ & \dots & $<0.5$ & \dots  \\
O & 2.643 & @1.10   & $<1.8$ & $>2.3$ & $<0.7$ & 5.50 & $<1.7 $ &\dots & $>2.4$ & \dots & $<0.7$ & \dots  \\
Q & 4.033 & @0.80   & $<2.6$ & $>2.4$ & $<1.0$ & 1.71 & $<1.0 $ &\dots & $>6.3$ & \dots & $<0.7$ & \dots  \\
T & 7.588 & @3.84   & $<7.3$ & $>1.7$ & $<2.4$ & 4.07 & $<6.0 $ &\dots & $>2.0$ & \dots & $<2.2$ & \dots \\
U & 4.343 & @2.15   & $<3.4$ & $>2.0$ & $<1.2$ & 1.78 & $<1.7 $ &$>0.16$ & $>4.0$ & $<42$ & $<0.9$ & $>0.3$ \\
\hline
\end{tabular}

\label{tab:47tucpsrlimits}
\end{center}
\end{table*}

The limits for the pulsar parameters obtained with this constant
maximum acceleration are presented in Table~\ref{tab:47tucpsrlimits}.
All characteristic ages are larger than 170\,Myr and all pulsars have
magnetic fields lower than $2.4 \times 10^{9}$\,Gauss. These values are
consistent with those measured for Galactic millisecond pulsars (e.g.,
Camilo, Thorsett \& Kulkarni 1994)\nocite{ctk94}.

Note that these limits are extremely conservative, and are independent
of any detailed modeling of the cluster. It is nevertheless clear that
even with such a crudely over-estimated $a_{l\,\rm max}(R_{\perp})$ we
can derive useful limits for the parameters of the pulsars.

By modeling the cluster and obtaining better constraints for $a_{l\,\rm
max}(R_{\perp})$, more stringent limits can be derived. For this
reason, we now compare the observed pulsar $\dot{P}/P$ with the maximum
accelerations along each line of sight calculated from the King model
(Fig.~\ref{fig:accel}).  The limits for the pulsar parameters derived
from this model are also presented in Table~\ref{tab:47tucpsrlimits}.
As expected, these are more constraining than those derived before.  In
particular, the characteristic ages for 47~Tuc~C and M exceed 3\,Gyr,
and a simple average of the lower limits on age for all pulsars yields
$\overline{\tau_c} > 0.9$\,Gyr, while the magnetic field strengths of
47~Tuc~J and M are less than $3\times10^8$\,Gauss, very close to the
lowest values observed among Galactic disk pulsars.  These values
depend, of course, on the particular cluster model used in calculating
the accelerations.

Limits for the characteristic ages and magnetic fields of the pulsars
in M15 have been derived by Anderson \nocite{and92} (1992), using a
mass model for the cluster based on the surface luminosity density and
assuming a constant mass-to-light ratio. The majority of pulsars in
that cluster have characteristic ages $\sim 10^{10}$\,yr \cite{and92},
but there are signs of a relatively recent burst in pulsar formation.
At least two pulsars have maximum characteristic ages $\sim 10^{8}$\,yr
and magnetic fields of about (10--20)$\times 10^9$\,Gauss.  This is
attributed to the ongoing core collapse in M15, leading to much
increased recycling of pulsars. In 47~Tuc, on the other hand, there are
no visible signs of such a recent burst in pulsar formation:  all the
known pulsars for which a timing solution has been obtained have
characteristic ages (probably much) greater than 170\,Myr, weak
magnetic fields, and very short rotational periods.

\section{Binary pulsar systems in 47~Tuc}
\label{sec:binaries}

The binary systems containing pulsars in 47~Tuc are divided into two
main groups, segregated by companion mass \cite{clf+00}.  The first is
composed of binaries with very short orbital periods (1.5--5.5\,hr) and
companion masses $\sim 0.03\,\rm M_{\odot}$. We have timing solutions
for three of these (47~Tuc~I, J and O).  The second group has orbital
periods in the range 0.4--2.3\,d and companion masses $\sim 0.2\,\rm
M_{\odot}$; we refer to this group as ``normal'' binaries.  Of these,
47~Tuc~E, H, Q, T and U have a timing solution at the moment.

Coherent timing solutions give extremely detailed orbital information,
and often allow the measurement of very small eccentricities.  Most
millisecond pulsar--white dwarf systems in the Galactic disk have very
low, but measurably significant eccentricities (e.g., Camilo
1999)\nocite{cam98}. These are thought to result from tidal
interactions between the neutron star and convective cells in the
envelope of its companion during the giant phase of its evolution, and
a relationship between orbital period and eccentricity can be derived
for such systems \cite{phi92b}.  The millisecond pulsar--white dwarf
systems in globular clusters often have larger eccentricities
than those in the disk of the Galaxy \cite{phi92b}.  These are thought
to result from interactions with other stars in the cluster:  the
denser the environment, the more likely one is to find a relatively
large degree of eccentricity in the system.

The five eccentricities measured for binary pulsars in 47~Tuc (for the
normal binaries 47~Tuc~E, H, Q, T and U; see Table~\ref{tab:binaries})
are all much larger than one would expect for similar systems in the
Galactic disk, and are therefore, presumably, a fossil remnant of
gravitational interactions between the binary systems and other cluster
stars.  Rasio \& Heggie (1995) \nocite{rh95} calculate the expected
value of this eccentricity for a system with a particular binary period
that has interacted with other stars in a region of known density for a
certain length of time.  Unfortunately we do not know the relevant
densities or interaction time scales for the pulsars observed in 47~Tuc
with any degree of certainty, possibly not even to an order of
magnitude.  Nevertheless, taking plausible estimates for density and
time scales, we derived some ``predicted'' eccentricities.  Those for
47~Tuc~E, Q, and T are within an order of magnitude of the observed
values, which is as close to agreement as we can expect to attain given
the uncertainties in input parameters.  The computed eccentricity for
47~Tuc~U is two orders of magnitude below the observed value, while
that for 47~Tuc~H is under-predicted by a factor of 1000.

One possible explanation for the unexpectedly large eccentricity of
47~Tuc~U, and possibly that of 47~Tuc~H, is that they have in the past
spent a considerable amount of time in a region with a much higher
stellar density than they are found in at present.  For instance, if
47~Tuc~L (an isolated pulsar) had spent $10^{10}$\,yr in its present
location while part of a 2.35-day binary, its computed eccentricity
would be 0.15.  Therefore, some of the pulsars we observe relatively far
from the centre of the cluster may have non-circular orbits in the
cluster potential which take them through higher-density regions
periodically, or may have been ejected from higher-density regions to
their present locations, leading to some of the large eccentricities
measured.

Alternatively, the high eccentricity of the 47~Tuc~H binary system, by
far the largest in the cluster, could indicate that it obtained its
present companion as an already-formed white dwarf through an exchange
interaction.  However, such exchanges tend to produce even higher
eccentricities \cite{phi92b}.

\begin{figure}
\setlength{\unitlength}{1in}
\begin{picture}(0,3.5)
\put(0,0){\includegraphics{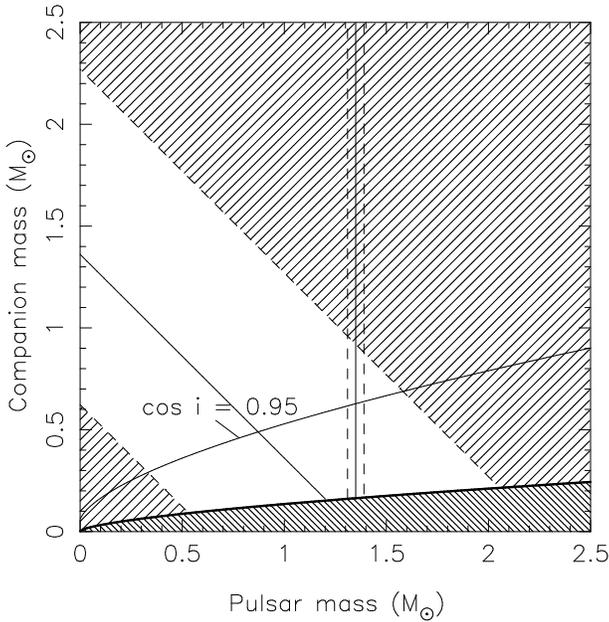}}
\end{picture}
\caption [] {Mass--mass diagram for the 47~Tuc~H system.  The allowed
range of total mass (sloping straight line surrounded by dashed lines)
is derived from the measured rate of advance of periastron and its
2-$\sigma$ uncertainty.  The area below the thick curving line is
excluded by the measured mass function and the requirement that $\cos i
\ge 0$.  There is a 95\% a priori probability that the system lies at
$\cos i \le 0.95$.  The vertical lines indicate the average measured
mass for neutron stars in the radio pulsar population, $1.35 \pm
0.04\,{\rm M}_{\odot}$ \protect\cite{tc99}. }

\label{fig:m1m2H}
\end{figure}

The large eccentricity of 47~Tuc~H has permitted a measurement of its
rate of advance of periastron: $\dot{\omega} = (0.059 \pm
0.024)^\circ\,\mbox{yr}^{-1}$ (Table~\ref{tab:binaries}).  Assuming
that this advance is entirely due to general relativity, and not to any
tidal effects (which is likely, since both stars are presumably
degenerate and have negligible dimensions compared to the orbital
separation), the total mass of the binary system is $1.4^{+0.9}_{-0.8}
{\rm M}_{\odot}$ (all uncertainties here are given at the 2-$\sigma$
level).  A mass-mass diagram is shown in Fig.~\ref{fig:m1m2H}, from
which limits on the individual masses can be derived: $m_p <
2.0\,\mbox{M}_{\odot}$ and $m_c > 0.1\,\mbox{M}_{\odot}$. It is also
improbable that $m_p < 0.4\,\mbox{M}_{\odot}$ or $m_c >
0.7\,\mbox{M}_{\odot}$, because this would require $\cos i > 0.95$.

\section{Conclusions and prospects}
\label{sec:concl}

In this paper we have presented phase-coherent timing solutions for 15
millisecond pulsars in 47~Tuc. Our main conclusions can be summarised
as follows.

All pulsars with known timing solutions are located less than
$1\farcm2$ from the cluster centre, and inside this region, their
spatial density is of the type $n(r) \propto r^{-2}$. This distribution
is different from that found in M15 ($n[r] \propto r^{-3.1}$, with no
well-defined outer limit), and possibly similar to the distribution of
blue stragglers in 47~Tuc with similar masses. Their confinement near
the centre of the cluster is real, and not a selection effect.  It is a
very important constraint on any model of the potential of 47~Tuc.

Two of the pulsars (47~Tuc~G and I) have a projected separation of only
600\,a.u., and similar $\dot{P}/P$.  Although it is unlikely, they may
belong to the first system known with two detectable pulsars, but
further timing measurements are required to investigate this.

We have measured the proper motions of five of the pulsars in 47~Tuc.
Only the general motion of the cluster is detectable, and the value
calculated from averaging the proper motions of the pulsars is
consistent with the proper motion of the cluster measured with {\em
Hipparcos\/}.

Lower limits for the surface mass density at several radial cuts near
the centre of the cluster were derived.  We have also used the values
of $(\dot{P}/P)_{\rm obs}$ to obtain limits on, or estimates of, the
accelerations at several projected distances from the centre of the
cluster.  These are consistent with a King model of the cluster using
accepted values for its distance to 47~Tuc, angular core radius and
central velocity dispersion.

We inferred upper limits for the magnetic fields of 15 pulsars: all are
below $2.4 \times 10^9$\,Gauss.  Also, no pulsar has a characteristic
age smaller than 170\,Myr.  Additionally we determined that the
characteristic ages of the pulsars are, on average, greater than
1\,Gyr.

The fact that none of the known binary pulsars has an orbital period
larger than 2.35 days is an important result. Simulations of their
formation \cite{rpr00} suggest that we should find binary pulsars with
orbital periods as large as about one month. However, despite the fact
that there are no selection effects against the detection of such
systems, we find none. This suggests that any binary pulsars with
longer periods were either hardened by interactions with other stars,
disrupted by the high stellar density near the centre of the cluster,
or were simply outnumbered by tight binaries formed at a later stage.
The timescale of such processes is proportional to the relaxation time,
i.e, it is proportional to the time between close encounters between
any two stars.

Binary formation and hardening is a source of kinetic energy for the
remaining stars in the cluster and tends to delay (or in some cases
reverse) core collapse \cite{spi87}; the lack of long-period binaries
therefore suggests that the core of 47~Tuc has had a density similar to
or larger than the present one for a time that is much longer than the
relaxation time, i.e., a few billion years \cite{hut92}.  If the
cluster had reached the present central density in the last few hundred
million years, and if it had never experienced a high-density phase
before, then it should still contain binary pulsars with long orbital
periods.  As we have also seen, the ages of the pulsars are of the
order of a few Gyr, and their large numbers in this cluster suggest
that vigorous millisecond pulsar formation occurred a long time ago,
which is consistent with the idea that the core of 47~Tuc was once, or
has been for a long time, in a high-density state.

Another piece of evidence of possible relevance to this question comes
from the higher than expected eccentricities of binaries like 47~Tuc~U
and H, the latter of which enabled us to measure its rate of advance of
periastron, yielding a total mass $m_p+m_c =
1.4^{+0.9}_{-0.8}$\,M$_{\odot}$.  These eccentricities suggest that the
product of stellar density and velocity dispersion near these binaries
was higher than is observed today for a long time, possibly several Gyr
(\S~\ref{sec:binaries}).  Whether this was so because the pulsars spent
a considerable amount of time nearer the present-density core, or
because the core was once denser, it would imply a dense core for a
very long period of time.

If the core of 47~Tuc was in fact relatively dense for a few Gyr, it is
likely that this contributed to the large numbers of
millisecond pulsars in the cluster (via the mechanisms of binary
formation and hardening in dense cores).  In that case the number of
millisecond pulsars in a cluster depends heavily on its previous
dynamical history. This could explain why clusters with similar
properties (core density, total mass, stellar mass distribution,
metallicity) have large differences in their pulsar populations, in
both number and kind \cite{ka96}. More complete searches for pulsars in
globular clusters, a better understanding of selection effects and a
good determination of the evolutionary states of globular clusters will
be needed to address this question.

A continuing timing programme will lead to additional and improved
measurements of the proper motions of the pulsars. A measurement of the
proper motions of the pulsars relative to the cluster would add
constraints to the cluster mass model.  Continued timing will also
allow a better determination of the orbital parameters of 47~Tuc~H, and
will test whether 47~Tuc~G and I are physically associated.  Additional
solutions for known pulsars may also be determined, and new pulsars
will almost certainly be discovered.  Continued monitoring of this
cluster is therefore likely to remain a worthy enterprise.

\section*{Acknowledgements}

We thank the skilled and dedicated telescope staff at Parkes for their
support during this project, Vicky Kaspi, Froney Crawford, Ingrid
Stairs and Jon Bell for assistance with observations, and Chris Salter,
Norbert Wex and David Nice for useful suggestions.  We are also
grateful to Justin Howell and colleagues for sharing their results with
us prior to publication, and to Fred Rasio for very useful comments
concerning the interpretation of the results. To Sterl Phinney we owe
the calculation of the King model of 47~Tuc used in this paper.  The
Parkes Telescope is part of the Australia Telescope which
is funded by the Commonwealth of Australia for
operation as a National Facility managed by CSIRO.  PCF gratefully
acknowledges support from Funda\c{c}\~{a}o para a Ci\^{e}ncia e a
Tecnologia through a Praxis~XXI fellowship under contract no.
BD/11446/97.  FC is supported by NASA grant NAG~5-9095.

\end{document}